\documentclass[12pt]{article}

\usepackage[a4paper,left=30mm,right=20mm,top=20mm,bottom=20mm]{geometry}
\usepackage{graphics}
\usepackage{amsmath}
\usepackage{amssymb}
\usepackage{epsfig}
\usepackage{cite}
\usepackage{xcolor}
\usepackage{ctable}
\usepackage{tabularx}
\usepackage[unicode]{hyperref}
\graphicspath{{/}}

\title{Prompt charmonia hadroproduction in High Energy Factorization with tree-level NLO corrections}

\author{A.A.~Prokhorov$^{1,2}$, S.P.~Baranov$^{3}$, A.V.~Lipatov$^{2}$, M.A.~Malyshev$^{2,4}$}

\begin{document}

\maketitle

\begin{center}

{\it $^{1}$Joint Institute for Nuclear Research, 141980 Dubna, Moscow region, Russia}\\
{\it $^{2}$Skobeltsyn Institute of Nuclear Physics, Lomonosov Moscow State University, 119991 Moscow, Russia}\\
{\it $^{3}$P.N.~Lebedev Institute of Physics, 119991 Moscow, Russia}\\

{\it $^{4}$Moscow Aviation Institute, 125993 Moscow, Russia}\\

\end{center}

\vspace{0.5cm}

\begin{center}

{\bf Abstract }

\end{center}

We consider prompt $J/\psi$, $\chi_{cJ}$ and $\eta_c$ production in high energy
$pp$ collisions in the framework of non-relativistic QCD. We apply high-energy
factorization with taking into account tree-level next-to-leading order
corrections to the production amplitudes.
Using several CCFM-evolved gluon densities in the proton and a previously
proposed matching scheme to merge the leading and higher-order contributions,
we extract the long-distance matrix elements for the entire charmonium family
from a simultaneous fit to available LHC data. We provide predictions
for the transverse momentum spectra and relative production rates
$\sigma(\psi^\prime)/\sigma(J/\psi)$ and $\sigma(\chi_{cJ})/\sigma(J/\psi)$
and find them to be in good agreement with the LHC data within the
theoretical and experimental uncertainties across a wide kinematic region.
We demonstrate that taking into account the tree-level NLO corrections improves
the global description of the data.

\vspace{1cm}


\newpage

\section{Introduction}\indent

The production of charmonia, the bound states of charmed quark-antiquark pairs,
is a long-standing mystery of modern high energy particle
physics~\cite{Quarkonia-Review-1, Quarkonia-Review-2, Quarkonia-Review-3}.
These processes provide a unique laboratory to investigate the interplay
between perturbative and non-perturbative phenomena within Quantum
Chromodynamics (QCD), as they involve the creation of charmed quark pairs at
high energy together with their subsequent soft transition into bound states.
The non-relativistic QCD (NRQCD)~\cite{NRQCD-1, NRQCD-2,NRQCD-3} is generally
considered as the most suitable theoretical tool to describe the formation and
decay of charmonium.
The NRQCD approach is a perturbative double-series expansion in the strong
coupling $\alpha_s$ and small relative velocity $v_c$ of the charmed quarks
in a bound state ($v_c^2 \sim 0.23$)~\cite{NRQCD-2,NRQCD-3}.
Its key feature is in systematically taking into account all possible Fock
states $n = |^{2S+1}L^{[a]}_J\rangle$ of a perturbatively produced
$c\bar c$ pair having spin $S$, orbital angular momentum $L$, total angular
momentum $J$ and color representation $a$ (singlet, $a = 1$, or octet, $a = 8$).
The non-perturbative transition from a given Fock-state into a physical
meson proceeds via soft gluon radiation and is encoded in the long distance
matrix elements (LDMEs).

Remarkable progress in the description of the $S$-wave and $P$-wave charmonia production
in $pp$, $p\bar p$, $ep$ and $\gamma p$
collisions has been achieved within the next-to-leading (NLO) calculations
(see, for example,~\cite{jpsi-psi-NRQCD-1, jpsi-psi-NRQCD-2, GlobalFit-psi2S,
chic-NRQCD-1, chic-NRQCD-2} and references therein), where corresponding
LDMEs were extracted from the fits to the world data.
However, there remain several well-known problems, such as the "polarization
puzzle"~\cite{PolarizationPuzzle-1, PolarizationPuzzle-2, PolarizationPuzzle-3}
and the "heavy quark spin symmetry puzzle"~\cite{HQSSPuzzle-1, HQSSPuzzle-2},
which still await a commonly accepted solution.

Making use of the high energy factorization~\cite{HighEnergyFactorization}
(or $k_T$-factorization~\cite{kt-factorization}) approach
can be considered as a convenient alternative to explicit high order pQCD
calculations as it effectively includes the leading-log part of higher order
corrections absorbing them into the Transverse Momentum-Dependent (TMD, or
unintegrated) gluon densities in a proton. The summation of the terms proportional to
$\alpha_s^n \ln^n s/\Lambda_{\rm QCD}^2 \sim \alpha_s^n \ln^n 1/x $ up to
infinitely large $n$ is made through the parton evolution equations:
Balitsky-Fadin-Kuraev-Lipatov (BFKL)~\cite{BFKL}
or Ciafaloni-Catani-Fiorani-Marchesini (CCFM)~\cite{CCFM}.
More details can be found in reviews~\cite{TMD-Review, TMD-Review2}.
It was already shown that, even at the leading-order (LO) perturbative
calculations, the incorporation of the $k_T$-factorization with
the NRQCD formalism improves the overall description of collider data,
see~\cite{Our-Charmonia-1, Our-Charmonia-2, Our-Charmonia-3}.

Further progress in the description of the $\chi_{cJ}$ and $\psi^\prime$
production properties has recently been made~\cite{Our-Chic-NLO, Our-psi2S-NLO},
where the tree-level NLO contributions were added to the respective off-shell
production amplitudes.
To accurately combine the LO and NLO contributions in order to avoid double
counting between them, a consistent merging procedure (denoted as the NLO$^{\dag}$ calculation scheme), was proposed and
validated~\cite{TMD-Review, TMD-Review2}.
Besides that, we have adopted a special gluon radiation
model~\cite{TransitionMechanism} to describe the formation of mesonic states.
In due time, this model was shown to provide a consistent solution to the
"polarization puzzle"~\cite{Our-Charmonia-1, Our-Charmonia-2, Our-Charmonia-3} (see also~\cite{Our-Bottomonia}).

In the present paper we apply the developed NLO$^{\dag}$ scheme~\cite{Our-Chic-NLO, Our-psi2S-NLO} to prompt $J/\psi$ and $\eta_c$ hadroproduction and also reconsider prompt $\chi_{cJ}$ hadroproduction. The latter is motivated mainly by the opportunity to incorporate into our fitting procedure for charmonia LDMEs some specific observables, such as the cross section ratios $\sigma(\chi_{cJ})/\sigma(J/\psi)$,
which were previously inaccessible without an analysis of $J/\psi$.
Our goal is to achieve a consistent, simultaneous description of all available unpolarized data on the entire charmonium family across the whole kinematic region with the use of high energy factorization. We wish to improve our previous LO calculations~\cite{Our-Charmonia-1, Our-Charmonia-2, Our-Charmonia-3}
by properly taking into account the tree-level NLO contributions.
Here we concentrate on the accurate determination of charmonia LDMEs
within the NLO$^{\dag}$ approach and on the description
of the measured transverse momentum distributions,
leaving the polarization observables
for a forthcoming dedicated study.

The paper is organized as follows. In Section~2, we briefly describe our
theoretical framework and basic steps of our calculations. In Section~3,
we present our numerical results and provide some discussions.
Our final conclusions are summarized in Section~4.

\section{Theoretical framework} \indent

Our present calculations follow the guideline of our previous
works~\cite{Our-Chic-NLO, Our-psi2S-NLO}. Here we only briefly recall
the most relevant steps.
At the LO in $v_c$, the relative velocity of the quarks,
the production of $\psi = J/\psi, \psi^{\prime}$ mesons proceeds solely via the color singlet
state $n = \, ^3S_1^{[1]}$ and is represented by the ${\cal O}(\alpha_s^3)$
off-shell gluon-gluon fusion subprocess
$g^* + g^* \to c\bar c \left[ ^3S_1^{[1]}\right] + g$.
At higher orders in $v_c$, the color octet states $n = \, ^1S_0^{[8]}$,
$^3S_1^{[8]}$ and $^3P_J^{[8]}$ with $J = 0$, $1$ and $2$ come into play.
The leading ${\cal O}(\alpha_s^2)$ contributions to the color octet (CO) production
are represented by the simplest $2\to 1$ subprocesses $g^* + g^* \to c\bar c [n]$, whereas the $2 \to 2$ subprocesses $g^* + g^* \to c\bar c[n] + g$ are of the order
${\cal O}(\alpha_s^3)$ and regarded as the NLO$^{\dag}$ corrections.
Likewise, in case of $\chi_{cJ}$ and $\eta_{c}$ production we consider the same $g^* + g^* \to c\bar c [n]$ and $g^* + g^* \to c\bar c[n] + g$ subprocesses with $n = \, ^3P_J^{[1]},\,^3S_1^{[8]}$ and $n = \, ^1S_0^{[1,8]},\,^3S_1^{[8]},\,^1P_1^{[8]}$, respectively.

In the high energy factorization, the interacting gluons have non-zero
virtualities (or, equivalently, non-zero transverse momenta) and longitudinal
components in their polarization vectors.
The necessary tree-level off-shell production amplitudes were derived in our
previous works~\cite{Our-Charmonia-1, Our-Chic-NLO}. The gauge invariance of
the relevant expressions was achieved by employing the effective QCD vertices
(see~\cite{EffectiveVertices} and references therein)
and has been explicitly tested.

It is important that
our CO production amplitudes incorporate the formation of bound states through soft gluon emissions described in terms of
chromo-electric dipole E1 transitions, as it
is proposed in a model~\cite{TransitionMechanism}
(see also~\cite{Our-Charmonia-1, Our-Charmonia-2, Our-Charmonia-3}
for more information). The probabilities to form each of the bound
states are determined by the corresponding LDMEs.

The differential cross sections of a considered processes are calculated as
a convolution of an off-shell production amplitude $\overline {|{\cal A}|^2}$ and TMD gluon densities
in the proton, $f_g(x, {\mathbf k}_T^2, \mu^2)$:

\begin{gather}
  \sigma_{2 \to 1} = \int{2\pi \over x_1\,x_2\,s\,F}\; f_g(x_1, {\mathbf k}_{1T}^2, \mu^2) f_g(x_2, {\mathbf k}_{2T}^2, \mu^2) \overline{|{\cal  A}_{2\to1}|^2} \, d{\mathbf k}_{1T}^2 d{\mathbf k}_{2T}^2 dy {d\phi_1 \over 2\pi} {d\phi_2 \over 2\pi}, \\
  \sigma_{2 \to 2} = \int{1\over 8\pi\,x_1\,x_2\,s\,F}\;f_g(x_1, {\mathbf k}_{1T}^2, \mu^2) f_g(x_2, {\mathbf k}_{2T}^2, \mu^2) \overline{|{\cal  A}_{2\to2}|^2} \times \nonumber \\
  \times d{\mathbf p}_{T}^2 d{\mathbf k}_{1T}^2 d{\mathbf k}_{2T}^2 dy dy_g {d\phi_1 \over 2\pi} {d\phi_2 \over 2\pi},
\end{gather}

However, when combining the LO and NLO contributions, one needs to take care
to avoid double counting (see, for example,\cite{TMD-Review}). Indeed, the emission of the same gluon can be
considered as part of the evolution of parton densities and as part of the
hard subprocess, and so, counted twice.
To separate the gluon emissions from the different origins, we have developed
an appropriate merging procedure~\cite{Our-Chic-NLO} for $2\to 1$ and
$2\to 2$ contributions.
The key idea is to introduce a special double-counting-exclusion (DCE) cut for $2 \to 2$ subprocess, which excludes the gluons already generated by the non-collinear parton evolution in the $2 \to 1$ subprocess and ensures that the hardest gluon emission in the $2 \to 2$ subprocess comes from the hard matrix elements
(see also~\cite{MatchingProcedure-1, MatchingProcedure-2}).
It was first proposed and successfully applied to the production of
$\chi_{cJ}$ mesons~\cite{Our-Chic-NLO}; then improved and extended to the region
of forward rapidities and low transverse momenta~\cite{Our-psi2S-NLO} for correct application within the LHCb acceptance. Our present calculations strictly follow
the developed approach~\cite{Our-Chic-NLO, Our-psi2S-NLO}.

\subsection{Numerical parameters} \indent
\label{sec:num_par}

In our numerical calculations, we employ two sets of the TMD gluon densities in a proton,
namely, A0~\cite{A0} and JH'2013 set 2~\cite{JH2013}. They are obtained from a
numerical solution of the Catani-Ciafaloni-Fiorani-Marchesini (CCFM)
evolution equation~\cite{CCFM}, which resums large logarithmic terms
proportional to $\alpha_s^n \ln^n 1/x$ and $\alpha_s^n \ln^n 1/(1 - x)$.
The parameters of the starting (low-scale) distributions were determined
from fits to HERA data on the proton structure functions $F_2(x, Q^2)$ and
$F_2^c(x, Q^2)$ at small $x$.
Both the A0 and JH'2013 set 2 gluon densities are widely used in
phenomenological applications\footnote{They
are available from a commonly known package \textsc{tmdlib}~\cite{TMDLib2},
which is a C$++$ library providing a framework and an interface to many
different parametrizations. They are also implemented in the Monte-Carlo
event generators \textsc{pegasus}~\cite{PEGASUS} and
\textsc{cascade}~\cite{CASCADE}.}
(see, for example, \cite{Motyka-photon, LMJ-PP, LM-Higgs, LLM-FL, LLM-photon}
and references therein).

We set the  meson mass to $m(\eta_{c}) = 2.984$~GeV, $m(J/\psi) = 3.097$~GeV, $m(\chi_{c1}) = 3.511$~GeV, $m(\chi_{c2}) = 3.556$~GeV, $m(\psi^{\prime}) = 3.686$~GeV and charm quark $m_c = 1.5$~GeV; branching fractions $Br(J/\psi \to \mu^+\mu^-) = (5.96\pm0.03)$\%, $Br(\psi^{\prime} \to J/\psi X) = (61.6\pm0.7)$\%, $Br(\chi_{c1} \to J/\psi\gamma) = (33.9\pm1.2)$\%, $Br(\chi_{c2} \to J/\psi\gamma) = (19.1\pm0.8)$\%\cite{PDG2026}.

We use the one-loop expression for the QCD coupling $\alpha_s$ with $n_f = 4$
quark flavours and $\Lambda_{\rm QCD} = 250$~MeV for A0 gluon, and the two-loop
expression with $n_f = 4$ and $\Lambda_{\rm QCD} = 200$~MeV for JH'2013 set 2
gluon distribution~\cite{A0, JH2013}.
Our default choice for the renormalization scale $\mu_R$ is the transverse
mass $m_T$ of the produced charmonium. The factorization scale $\mu_F$ is
set to $\mu^2_F = \hat s + {\mathbf Q}_T^2$, where ${\mathbf Q}_T$
is the net transverse momentum of the initial off-shell gluon pair.
This choice of $\mu_F$ is specific for CCFM evolution (see~\cite{CCFM} for
more details).
However, in accordance with the proposed merging LO $+$ NLO
scheme~\cite{Our-psi2S-NLO, Our-Chic-NLO}, we shift the scale as
$\mu_F \rightarrow \mu_F = m_T$ for $2 \to 2$ color octet subprocesses.

\subsection{Charmonia wave functions, LDMEs and HQSS relations} \indent

\vskip 2mm
For the reader's convenience, here we briefly recall the definitions and normalization of the LDMEs used both in this study and our previous works.
For  $S$- and $P$-wave color singlet (CS) states we set:
\begin{gather}
\langle\mathcal{O}[^{2S+1}S^{[1]}_{J}]\rangle\ = 2N_{c}(2J+1)\frac{|\mathcal{R}(0)|^2}{4\pi}, \\
\langle\mathcal{O}[^{2S+1}P^{[1]}_{J}]\rangle\ = 6N_{c}(2J+1)\frac{|\mathcal{R}^{\prime}(0)|^2}{4\pi}
\end{gather}
where $|\mathcal{R}(0)|^2$ and $|\mathcal{R}^{\prime}(0)|^2$ are the squared CS wave function at the origin and its derivative, respectively. Such definition of color singlet LDMEs satisfy the Heavy Quark Spin Symmetry (HQSS). The CS LDMEs can be calculated within potential models or extracted from known decay widths.

In contrast, values of LDME's for CO intermediate states are free parameters not
predictable by the theory and have to be found from the experimental data.
The HQSS symmetry dictates the {following: 
\begin{gather}
\langle\mathcal{O}^{\psi}[^3P^{[8]}_{J}]\rangle = (2J+1)\langle\mathcal{O}^{\psi}[^3P^{[8]}_{0}]\rangle,\\
\label{eq:HQSS_3PJ}
\langle\mathcal{O}^{\chi_{cJ}}[^3S^{[8]}_{1}]\rangle = (2J+1)\langle\mathcal{O}^{\chi_{c0}}[^3S^{[8]}_{1}]\rangle, \\
\langle\mathcal{O}^{\eta_{c}}[^1P^{[8]}_{1}]\rangle = 3\langle\mathcal{O}^{J/\psi}[^3P^{[8]}_{0}]\rangle, \\
\langle\mathcal{O}^{\eta_{c}}[^1S^{[8]}_{0}]\rangle = \frac{1}{3}\langle\mathcal{O}^{J/\psi}[^3S^{[8]}_{1}]\rangle, \\
\langle\mathcal{O}^{\eta_{c}}[^3S^{[8]}_{1}]\rangle = \langle\mathcal{O}^{J/\psi}[^1S^{[8]}_{0}]\rangle.
\end{gather}
\noindent
The violation of the HQSS  relations primarily arises from spin-dependent terms in the effective NRQCD Lagrangian and scales as $\mathcal{O}(v^2)$. Nevertheless, we retain these relations in our analysis to reduce the number of independent LDMEs involved in the fitting procedure.

\subsection{Fitting procedure for $J/\psi$ and $\chi_{cJ}$ LDMEs}\indent
\label{sec:fit_procedure}

To determine the LDMEs for $J/\psi$ mesons,
we perform a global fit which includes all of the available
unpolarized LHC data on inclusive $J/\psi$ production. The initial concept for the fitting procedure strictly follows the methodology previously reported~\cite{Our-psi2S-NLO, Our-Chic-NLO}.

First, we extract the value of a CS LDME using well-known NLO expression for leptonic decay width $\Gamma(J/\psi\to e^{+}e^{-})$~\cite{Braaten_Lee}. The world average value $\Gamma(J/\psi\to e^{-}e^{+}) = 5.52$~keV~\cite{PDG2026} leads to  $\langle\mathcal{O}^{J/\psi}[^3S^{[1]}_{1}]\rangle = 1.16$ GeV$^{3}$. The fitting procedure for CO LDMEs involves $3$ unknown parameters, namely, $\langle\mathcal{O}^{J/\psi}[^1S^{[8]}_{0}]\rangle, \langle\mathcal{O}^{J/\psi}[^3P^{[8]}_{0}]\rangle $
and $\langle\mathcal{O}^{J/\psi}[^3S^{[8]}_{1}]\rangle $, while the other CO LDME's can be obtained from HQSS relation (\ref{eq:HQSS_3PJ}).

 According to the NRQCD formalism, the $J/\psi$
production cross section reads:
\begin{gather}
\frac{d\sigma^{J/\psi}}{dp_{T}} = \frac{d\sigma(c\bar{c}[^3S^{[1]}_1])}{dp_{T}}\langle\mathcal{O}^{J/\psi}[^3S^{[1]}_{1}]\rangle + \frac{d\sigma(c\bar{c}[^1S^{[8]}_0])}{dp_{T}}\langle\mathcal{O}^{J/\psi}[^1S^{[8]}_{0}]\rangle + \nonumber \\
 + \sum_{J}(2J+1)\frac{d\sigma(c\bar{c}[^3P^{[8]}_J])}{dp_{T}}\langle\mathcal{O}^{J/\psi}[^3P^{[8]}_{0}]\rangle +\frac{d\sigma(c\bar{c}[^3S^{[8]}_1])}{dp_{T}}\langle\mathcal{O}^{J/\psi}[^3S^{[8]}_{1}]\rangle + \nonumber \\
 + \frac{d\sigma^{\chi_{cJ}\rightarrow J/\psi\gamma}}{dp_{T}} + \frac{d\sigma^{\psi^{\prime}\rightarrow J/\psi X}}{dp_{T}}.
\end{gather}
\noindent
The contributions from radiative decays of $\chi_{cJ}$ and $\psi^{\prime}$ could be calculated with LDME's obtained earlier in~\cite{Our-Chic-NLO, Our-psi2S-NLO}.

The values of $J/\psi$ LDMEs are determined by minimizing the $\chi^2$ function running
over all data points. As in case of $\psi^{\prime}$ production, we found that fitted LDMEs for $J/\psi$  mesons show strong negative correlation between
$^1S^{[8]}_{0}$ and $^3P^{[8]}_{J}$ contributions for both considered TMD gluon densities.
This makes it impossible to extract these parameters with acceptable accuracy. Therefore, we replace the individual parameters with a linear combination:
\begin{gather}
  \langle\mathcal{O}^{J/\psi}_{l.c.}\rangle = \langle\mathcal{O}^{J/\psi}[^3P^{[8]}_{0}]\rangle + r \times \langle\mathcal{O}^{J/\psi}[^1S^{[8]}_{0}]\rangle,
\label{eq:linear_comb_jpsi}
\end{gather}
 \noindent
where $r$ is defined as
\begin{gather}
r = \frac{d\sigma(c\bar{c}[^1S^{[8]}_0])/dp_{T}}{\sum (2J+1)d\sigma(c\bar{c}[^3P^{[8]}_J])/dp_{T}}.
\label{eq:linear_comb_jpsi_r}
\end{gather}
\noindent
The coefficient $r$ was calculated for every data point.
Its values were averaged over the whole rapidity range and, separately, over the
central and forward rapidity regions\footnote{The differences in the values of $r$
in the central and forward regions reflect only differences in the normalization
in the different regions of the phase space, while the shapes of the transverse
momentum distributions remain the same. This does not affect the obtained results
but can be useful for upcoming studies.}.
The resulting $r$ with a standard deviation are
collected in Table~\ref{tab:r_ratio} for both A0 and JH'2013 set 2 gluon distributions.

\begin{table}[ht]
\centering
\small
\caption{Values of $r$ calculated in different rapidity regions for A0 and JH'2013 set 2 gluon densities in a proton.}
\label{tab:r_ratio}
\vspace{0.2cm}
\begin{tabularx}{\textwidth}{l >{\centering\arraybackslash}X >{\centering\arraybackslash}X >{\centering\arraybackslash}X}
\toprule
TMD & Combined & Central & Forward \\
\midrule
\addlinespace[0.2cm]
A0 & $0.39 \pm 0.05$ & $0.36 \pm 0.02$ & $0.46 \pm 0.05$ \\
\addlinespace[0.2cm]
JH'2013 set 2 & $0.40 \pm 0.03$ & $0.39 \pm 0.03$ & $0.46 \pm 0.03$ \\
\addlinespace[0.2cm]
\bottomrule
\end{tabularx}
\end{table}

Additionally, we found that the global fit to prompt $J/\psi$ data is insensitive to separating the contributions from the $\sum\, ^3P_J^{[8]}$ intermediate states to direct $J/\psi$ production and from the $\sum ^3P_J^{[1]}$ states to radiative production $\chi_{cJ} \to J/\psi\,\gamma$. Likewise,
it is insensitive to separating the respective contributions in the $^3S_1^{[8]}$ channel. This insensitivity allows us to establish a linear regression between the corresponding LDMEs:
\begin{gather}
\langle\mathcal{O}^{J/\psi}_{l.c.}\rangle = a_1 - b_1\times|\mathcal{R}^{\prime\chi_{cJ}}|^2,
\label{eq:linear_comb_Jpsi_chic1} \\
\langle\mathcal{O}^{J/\psi}[^3S^{[8]}_1]\rangle = a_2 - b_2\times\langle\mathcal{O}^{\chi_{c0}}[^3S^{[8]}_1]\rangle.
\label{eq:linear_comb_Jpsi_chic2}
\end{gather}
\noindent
Our calculation shows that the coefficients $a_1$, $a_2$, $b_1$ and $b_2$ could be extracted with a very good accuracy for both A0 and JH'2013 set 2 gluon distributions.

This observation motivated us to perform a more comprehensive fitting procedure,
which also includes a revision of LDMEs for $\chi _{cJ}$ mesons. In fact, our previous 
fit~\cite{Our-Chic-NLO} include only the limited datasets, namely, the differential spectra $d\sigma (\chi_{cJ})/dp_T$ measured by ATLAS Collaboration at $\sqrt{s}=7$~TeV~\cite{dataset_chic_ATLAS7} and $d\sigma (\chi_{c1}+\chi_{c2})/dp_T$ measured by CDF at $\sqrt{s}=1.8$~TeV~\cite{dataset_chic_CDF} with $21$ points overall. Furthermore, the cross section ratio $R_{\chi_{cJ}/J/\psi} = \sigma (\chi_{cJ}\rightarrow J/\psi\gamma)/\sigma(J/\psi)$ was not available for analysis~\cite{Our-Chic-NLO}, since the LDMEs for $J/\psi$ mesons had not yet been determined at that time. In the present study, however, we are now able to include such observables and perform this comprehensive analysis.

Let us introduce our two-step fitting procedure\footnote{It should be noted that a single-step global fit with four parameters is ineffective. Due to the vast statistical dominance of the prompt $J/\psi$ data over the $\chi _{cJ}$ samples, the $\chi^2$ minimization would be severely biased, which could lead to an incorrect description of the $\chi _{cJ}$ data.} aimed at determining both the CO LDMEs  ($\langle\mathcal{O}^{J/\psi}_{l.c.}\rangle$, $\langle\mathcal{O}^{J/\psi}[^3S^{[8]}_1]$)
for $J/\psi$ mesons and $|\mathcal{R}^{\prime\chi_{cJ}}(0)|^2$, $\langle\mathcal{O}^{\chi_{c0}}[^3S^{[8]}_1]\rangle$ for $\chi_{cJ}$ mesons. We highlight that the relations $\eqref{eq:linear_comb_Jpsi_chic1}-\eqref{eq:linear_comb_Jpsi_chic2}$ allow us to keep just 2 independent free parameters.
\begin{itemize}
\item[$\text{\textcircled{\small 1}}$] At the first step, we utilize all available data of differential cross section $d\sigma(J/\psi)/dp_T$ on prompt $J/\psi$ production to extract the numerical coefficients of the linear combinations $\eqref{eq:linear_comb_Jpsi_chic1}-\eqref{eq:linear_comb_Jpsi_chic2}$. The full list of the used datasets are presented in Table~\ref{tab:Jpsi_chi2}.

\item[$\text{\textcircled{\small 2}}$] At the second step, we use the observables that include the direct measurements of $\chi_{cJ}$ production. The full list of the used datasets are presented in Table~\ref{tab:ChicJ_chi2}. The values $|\mathcal{R}^{\prime\chi_{cJ}}(0)|^2$ and $ \langle\mathcal{O}^{\chi_{c0}}[^3S^{[8]}_1]\rangle$ are determined by minimizing the $\chi^2$ function ovel all data points.  After that, the LDMEs for $J/\psi$ mesons are calculated from $\eqref{eq:linear_comb_Jpsi_chic1}-\eqref{eq:linear_comb_Jpsi_chic2}$.
\end{itemize}

\noindent
 It is important that fitting results for $\chi _{cJ}$ are sensitive to experimental uncertainties related to spin alignment. As it is known, the LHC data~\cite{CMS_polarization_chi_cJ} indicate opposite signs of the $\lambda _{\theta }$ polarization parameters  for $\chi _{c1}$ and $\chi _{c2}$ mesons\footnote{Our previous calculations~\cite{Our-Chic-NLO} were consistent with this assessment.}. To avoid artificially overestimating or underestimating the experimental errors, the following rule was applied: for observables representing the sum of the $\chi_{c1} + \chi_{c2}$ contributions, the spin-alignment uncertainty was omitted (as these uncertainties are expected to largely compensate for each other); in all other cases, this uncertainty was explicitly taken into account.

Note that the comparison of our previous calculations~\cite{UnEqualWaveFunctions, Our-Charmonia-2, Our-Charmonia-3, Our-Chic-NLO} with available experimental data for $\sigma (\chi_{c2})/\sigma (\chi_{c1})$ ratio points to non-equality of $|\mathcal{R}^{\prime\chi_{c1}}(0)|^2$ and $|\mathcal{R}^{\prime\chi_{c2}}(0)|^2$.
Such an effect can qualitatively be attributed to spin-orbit and tensor interactions in the $P$-wave states, though a rigorous quantitative NRQCD calculation is still missing. Moreover, the values $|\mathcal{R}^{\prime\chi_{c0}}(0)|^2 = 0.0408$~GeV$^5$ and $|\mathcal{R}^{\prime\chi_{c2}}(0)|^2 = 0.0575$~GeV$^5$ extracted from the decay widths $\Gamma({\chi_{c0}\to \gamma\gamma}) = 2.43$~keV and $\Gamma({\chi_{c2}\to \gamma\gamma}) = 0.56$~keV also point to a significant violation of the HQSS relations. Therefore, we extend our fitting procedure to investigate these effects.
Specifically, we consider an alternative scenario for the second step, where the parameters $|\mathcal{R}^{\prime\chi_{c1}}(0)|^2$ and $|\mathcal{R}^{\prime\chi_{c2}}(0)|^2$ are allowed to vary independently. For relation \eqref{eq:linear_comb_Jpsi_chic1} we use  "averaged" $|\mathcal{R}^{\prime\chi_{cJ}}(0)|^2$ calculated as:
\begin{gather}
|\mathcal{R}^{\prime\chi_{cJ}}(0)|^2\times \frac{d\sigma (c\bar{c}[^3P^{[1]}_1]) + c\bar{c}([^3P^{[1]}_2])}{dp_T} = \nonumber \\ = |\mathcal{R}^{\prime\chi_{c1}}(0)|^2\times \frac{d\sigma (c\bar{c}[^3P^{[1]}_1])}{dp_T} + |\mathcal{R}^{\prime\chi_{c2}}(0)|^2\times \frac{d\sigma (c\bar{c}[^3P^{[1]}_2])}{dp_T}.
\end{gather}
\noindent
The results of two-step fitting procedure  for both A0 and JH'2013 set 2 gluon densities
are presented in the next section. The results with unequal $|\mathcal{R}^{\prime\chi_{c1}}(0)|^2$ and $|\mathcal{R}^{\prime\chi_{c2}}(0)|^2$ are marked with the label 'neq. $|\mathcal{R}|^2$' throughout the text, tables and figures.

\subsection{LDMEs for $\eta_{c}$} %\label{sec:etac_LDME}
\indent

Studying the $\eta_{c}$ production is of particular interest as it allows us to test the theoretical predictions using the HQSS relations discussed above. Since all the relevant CO LDMEs are strictly linked to those of the $J/\psi$ meson, it allows a parameter-free calculation of the $\eta _{c}$ cross sections.

{As before, the CS LDME is extracted from the $\eta_{c}\rightarrow\gamma\gamma$ decay width using the corresponding
expression derived in the NLO approximation~\cite{Braaten_Lee}. However, the experimental measurements of the decay width vary significantly. For instance, the average value reported in 2024 by the Particle Data Group (PDG) Collaboration
is $\Gamma({\eta_{c}\to\gamma\gamma}) = 5.063\pm0.405$~keV~\cite{PDG2024}. The latest BES III experimental
measurement is $\Gamma({\eta_{c}\to\gamma\gamma}) = 7.48\pm1.51$~keV~\cite{BESIII_etac_decay}. The
updated PDG average width is $\Gamma({\eta_{c}\to\gamma\gamma}) = 6.43\pm0.46$ keV~\cite{PDG2026}. Below we adopt $\langle\mathcal{O}^{\eta_{c}}(^1S^{[1]}_0)\rangle = 0.28 \pm 0.05$ GeV$^3$, where the central value corresponds to the latest PDG value and uncertainties cover the decay width variations listed above.

The extracted CS LDMEs for $\eta_c$ and $J/\psi$ mesons are consistent with the HQSS relation.  The visible discrepancy can be attributed to relativistic ${\cal O}(v_c^2)$ corrections, the dominant one for $S$-wave states being the hyperfine splitting.

\section{Numerical results} \indent

\begin{figure}
\begin{center}
%{\includegraphics[width=.5\textwidth]{Chi2_graph.pdf}}\hfill
{\includegraphics[width=.5\textwidth]{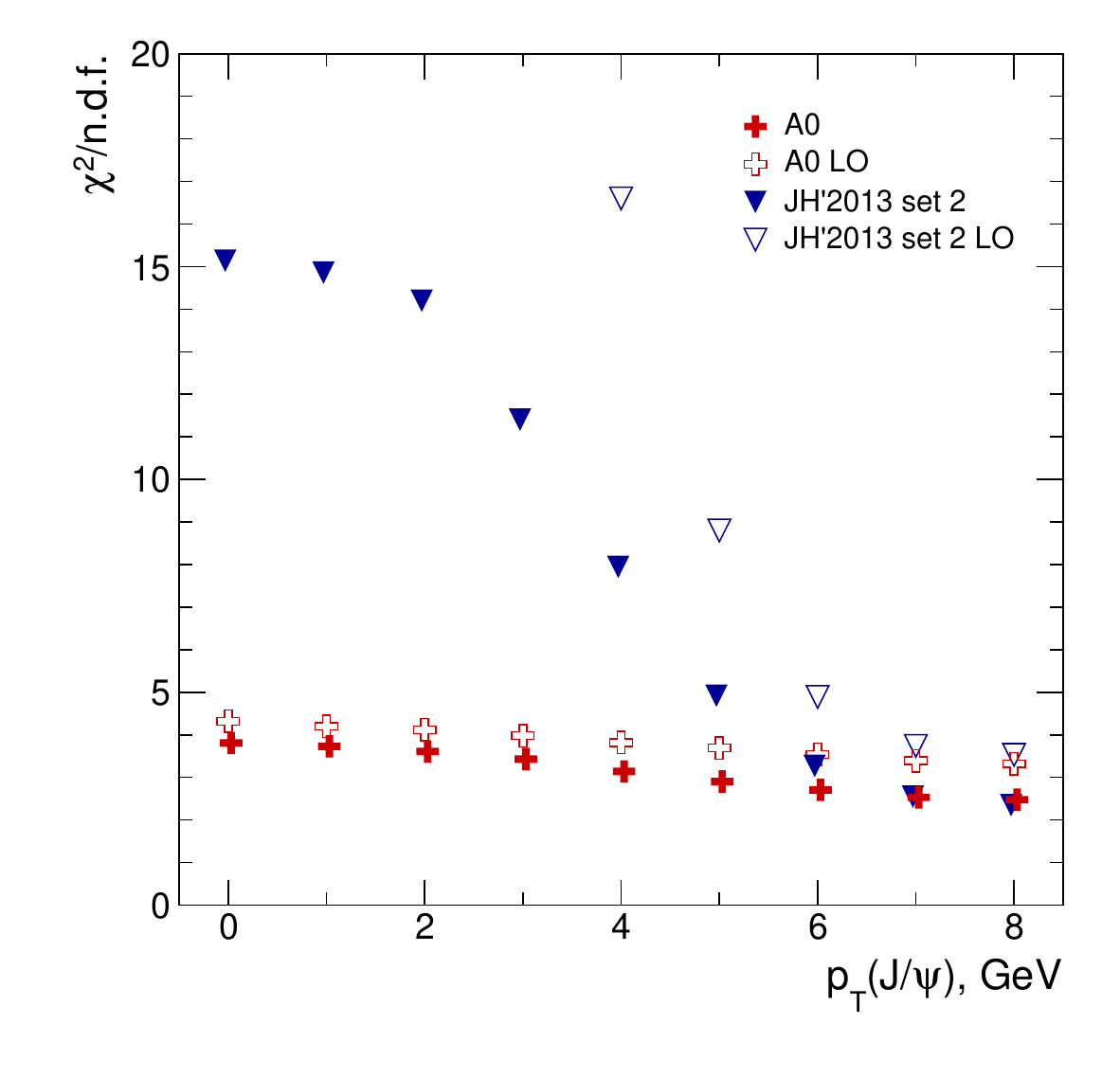}}\hfill
\caption{Lowest values of $\chi^2/{\rm n.d.f.}$ as function of minimal $J/\psi$  transverse momenta.}
\label{fig:Chi2_graph}
 \end{center}
\end{figure}

In this section, we present the results of our global two-step fitting. In Fig.~\ref{fig:Chi2_graph} we show the lowest values of $\chi^2/{\rm n.d.f.}$
obtained from the combined fit to all prompt $J/\psi$ differential cross section ($d\sigma/dp_T$) data  for different cutoffs $p_T^{\rm min}$ on the quarkonium transverse momentum. The full list of included data is presented in Table~\ref{tab:Jpsi_chi2}. The resulting distributions exhibit the same behavior as we found earlier in the global fit to the $\psi^{\prime}$ data~\cite{Our-psi2S-NLO}, leading to the same conclusions.
Specifically, for the A0 gluon density, an accurate simultaneous description of the entire LHC dataset is achieved regardless of the $p_{T}$ constraint. In case of JH'2013 set 2 gluon density, the results closely agree with A0 predictions only when the $p_T < 6$ GeV region (containing the forward rapidity data) is excluded. This constraint is identical to the cutoff introduced previously in the fit to the $\psi^{\prime}$ data, see~\cite{Our-psi2S-NLO}. This fact underscores that the observed discrepancy in the $p_T$ spectra is primarily driven by the distinct dynamics of the TMD gluon distributions at low ${\mathbf k}_T^2$ (see~\cite{A0,JH2013}). Besides that, the inclusion of tree-level NLO corrections allows us to improve the overall description of data in comparison with pure LO, especially for JH'2013 set 2 gluons distribution. At the same time, we note a general growth of minimal $\chi^2$ values compared to the case of $\psi^{\prime}$, which is mainly caused by the significantly reduced experimental uncertainties of $J/\psi$ datasets.

\begin{table}[ht]
\centering
\small
\caption{The values of LDMEs for $J/\psi$ meson fitted with available gluon densities in a proton.}
\label{tab:LDME}
\vspace{0.2cm}
\begin{tabularx}{\textwidth}{l >{\centering\arraybackslash}X >{\centering\arraybackslash}X >{\centering\arraybackslash}X >{\centering\arraybackslash}X}
\toprule
LDME & A0 & JH'2013 set 2 & A0 neq. $|R|^2$ & JH'2013 set 2 neq. $|R|^2$ \\
\midrule
\addlinespace[0.2cm]
$|\mathcal{R}^{\prime\chi_{c1}}(0)|^2\times 10/\mathrm{GeV}^{5}$
 & $0.375\pm0.014$ & $0.578\pm0.027$  & $0.593\pm0.027$ & $0.865\pm 0.038$ \\
\addlinespace[0.2cm]
$|\mathcal{R}^{\prime\chi_{c2}}(0)|^2\times 10/\mathrm{GeV}^{5}$
 & $0.375\pm0.014$ & $0.578\pm0.027$  & $0.249\pm0.014 $ & $0.337\pm0.026$ \\
\addlinespace[0.2cm]

$\langle\mathcal{O}^{\chi_{c0}}[^3S_1^{}]\rangle\times 10^{4}/\mathrm{GeV}^{3}$
 & $4.58\pm0.41$ & $3.83\pm0.43$  & $3.09\pm0.35$ & $3.01\pm0.39$ \\
\specialrule{0.2pt}{0.15cm}{0.15cm}
$\langle\mathcal{O}^{J/\psi}_{l.c.}\rangle\times 10^{3}/\mathrm{GeV}^{5}$
 & $8.54\pm0.12$ & $14.6\pm0.2$ & $8.12\pm0.17$ & $14.2\pm0.2$ \\
\addlinespace[0.2cm]

$\langle\mathcal{O}^{J/\psi}[^3S_1^{[8]}]\rangle\times 10^{4}/\mathrm{GeV}^{3}$
 & $10.8\pm0.6$ & $5.49\pm0.52$  & $12.8\pm0.5$ & $6.48\pm0.47$ \\
\addlinespace[0.2cm]
\bottomrule
\end{tabularx}
\end{table}

In this regard, we will further compare the fitting results obtained with the A0 and JH'2013 set 2 gluon functions, enforcing the $p_T > 6$ GeV constraint for the latter density. 
 Table~\ref{tab:LDME} presents the LDMEs obtained from the two-step fitting procedure described in Sec.~\ref{sec:fit_procedure}.
The experimental uncertainties and uncertainties in the branching fractions of the $\chi_{cJ} \to J/\psi + \gamma$ and $\psi^{\prime} \to J/\psi + X$ decays were taken into account during the fitting.
The results, along with their standard deviations 
are given for both scenarios: with identical and independent values of $|\mathcal{R}^{\prime\chi_{cJ}}(0)|^2$. As can be inferred from the table, the extracted LDMEs for $J/\psi$ remain fairly close within the uncertainties for both scenarios, whereas the derivative of CS wave function for $\chi_{cJ}$ mesons significantly deviates from its ‘average’ value and violates the HQSS relation. Specifically, the deviations for $|\mathcal{R}^{\prime\chi_{c1}}(0)|^2$ and $|\mathcal{R}^{\prime\chi_{c2}}(0)|^2$ reach approximately 50\% in opposite directions, which enables an excellent description of the $\sigma(\chi_{c2})/\sigma(\chi_{c1})$ ratio. Unfortunately, these numerical results indicate corrections well beyond the expected scales of NRQCD relativistic effects. On the other hand, the incorporation of the tree-level NLO$^{\dag}$ corrections allows us to reduce the difference between $|\mathcal{R}^{\prime\chi_{c1}}(0)|^2$ and $|\mathcal{R}^{\prime\chi_{c2}}(0)|^2$ by up to a factor of two in comparison with LO case\footnote{Violation of the HQSS relations for $\chi_{cJ}$ meson CS wave functions could be also significantly reduced in the specific and more complicated LO + NLO merging scenario, see\cite{Our-Chic-NLO}.} (see~\cite{Our-Charmonia-3} and Table~1 therein). Note that special assumption on the violation of HQSS relations for $\langle\mathcal{O}^{\chi_{cJ}}[^3S^{[8]}_{1}]\rangle$ leads to improvement of the LO results~\cite{Our-Charmonia-3}.
However, to avoid introducing additional free parameter, this effect is not considered in the present study. Nevertheless, further investigation into this issue remains a subject of great interest for future studies.

\begin{table}[ht]
\centering
\small
\caption{${\chi}^{2}/\mathrm{n.d.f.}$ for available $\chi_{cJ}$ datasets at $\mathrm{LO}+\mathrm{NLO}^{\dag}$.}
\label{tab:ChicJ_chi2}
\vspace{0.2cm}
\begin{tabularx}{\textwidth}{l l c >{\centering\arraybackslash}X >{\centering\arraybackslash}X >{\centering\arraybackslash}X >{\centering\arraybackslash}X}
\toprule
Observable & Experiment & n.d.f. & A0 & JH'2013 set 2 & A0 neq. $|R|^2$ & JH'2013 set 2 neq. $|R|^2$ \\
\midrule
% --- Величина 1 ---
$\mathrm{d}\sigma(\chi_{c1})/\mathrm{d}p_T$
 & ATLAS 7 TeV {\scriptsize \cite{dataset_chic_ATLAS7}}  & 5    & 3.55 & 2.76 & 1.99 & 0.82 \\
\specialrule{0.2pt}{0.15cm}{0.15cm}
% --- Величина 2 ---
$\mathrm{d}\sigma(\chi_{c2})/\mathrm{d}p_T$
 & ATLAS 7 TeV {\scriptsize \cite{dataset_chic_ATLAS7}} & 5   & 1.29 & 1.08 & 0.07 & 0.16 \\
\specialrule{0.2pt}{0.15cm}{0.15cm}
% --- Величина 3 ---
$\mathrm{d}\sigma(\chi_{cJ})/\mathrm{d}p_T$
 & CDF 1.8 TeV {\scriptsize \cite{dataset_chic_CDF}}  & 11(9)   & 0.53 & 0.09 & 0.24 & 0.06 \\
\specialrule{0.2pt}{0.15cm}{0.15cm}
% --- Величина 4 ---
$R_{\chi_{cJ}/J/\psi}$
 & ATLAS 7 TeV {\scriptsize \cite{dataset_chic_ATLAS7}}  & 5    & 0.22 & 0.15 & 0.65 & 0.21 \\
 & LHCb 7 TeV {\scriptsize \cite{dataset_chic_LHCb2}}     & 12(8) & 0.11 & 0.09 & 0.52 & 0.26 \\
\specialrule{0.2pt}{0.15cm}{0.15cm}
% --- Величина 5 (без первых двух колонок) ---
$R_{\chi_{c2}/\chi_{c1}}$
 & ATLAS 7 TeV {\scriptsize \cite{dataset_chic_ATLAS7}}     & 5    &      &      & 0.45 & 0.17 \\
 & CMS 7 TeV {\scriptsize \cite{dataset_chic_CMS}}      & 6    &      &      & 1.94 & 0.84 \\
 & LHCb 7 TeV {\scriptsize \cite{dataset_chic_LHCb}}    & 10(7)  &      &      & 1.62 & 0.21 \\
\midrule
\multicolumn{2}{l}{\textbf{Summary}}    & \textbf{59(50)} & \textbf{0.85} & \textbf{0.67} & \textbf{0.88} & \textbf{0.32}\\
\bottomrule
\end{tabularx}
\end{table}

Table~\ref{tab:ChicJ_chi2} summarizes the $\chi^2/n.d.f.$ evaluation results for all available experimental data on prompt $\chi_{cJ}$ meson production, specifying the observables, experiments and energies. The values in parentheses in the "n.d.f." column indicate the number of remaining data points after applying the cutoff $p_T >$ 6 GeV for JH'2013 set 2 gluon density. These calculations are performed using the LDMEs listed in Table~\ref{tab:LDME}. Since the scenario with identical $|\mathcal{R}^{\prime\chi_{cJ}}(0)|^2$ fails to reproduce the ratio $\sigma(\chi_{c1})/\sigma(\chi_{c2})$, these observables were not included in the $\chi_{cJ}$ fitting procedure in the given scenario, in order to maintain a physically consistent result\footnote{Naturally, these observables were also excluded when extracting the LDMEs in the scenario with identical $|\mathcal{R}^{\prime\chi_{cJ}}(0)|^2$.}.

The calculated observables are presented in Figs.~\ref{fig:ATLAS_CDF_chicJ} -- \ref{fig:ATLAS_LHCb_chicJ_Jpsi} for both the A0 and JH'2013 set 2 gluon densities.
  Our theoretical predictions (indicated by solid lines for the identical $|\mathcal{R}^{\prime\chi_{cJ}}(0)|^2$ scenario and by dashed lines for the independent one) show good agreement with the experimental differential cross sections and $\sigma(\chi_{cJ})/\sigma(J/\psi)$ ratios across various kinematic regions. However, as demonstrated in Fig.~4, the $\sigma(\chi_{c2})/\sigma(\chi_{c1})$ ratio can be simultaneously described well in both the central and forward rapidity regions only when independent values of $|\mathcal{R}^{\prime\chi_{cJ}}(0)|^2$ are employed.

	The theoretical uncertainty bands (shaded areas) are shown only for the A0 gluon density. Here, the inner (pink) bands represent the variation in the energy of soft gluons emitted during the color-octet transitions to the charmonium
bound state. Such uncertainties arise within the specific model~\cite{TransitionMechanism} applied to describe the $c\bar c$ bound state formation. Here we estimate them in exactly the same way as was done earlier~\cite{Our-psi2S-NLO, Our-Chic-NLO}.
    The outer (orange) bands include the renormalization scale uncertainties added in quadrature. As is customary~\cite{A0}, variations in the renormalization scale are accompanied by the simultaneous use of the $\text{A}0-$ and $\text{A}0+$ gluon densities for the scale choices $\mu_R/2$ and $2\mu_R$, respectively.

\begin{figure}
\begin{center}
{\includegraphics[width=.49\textwidth]{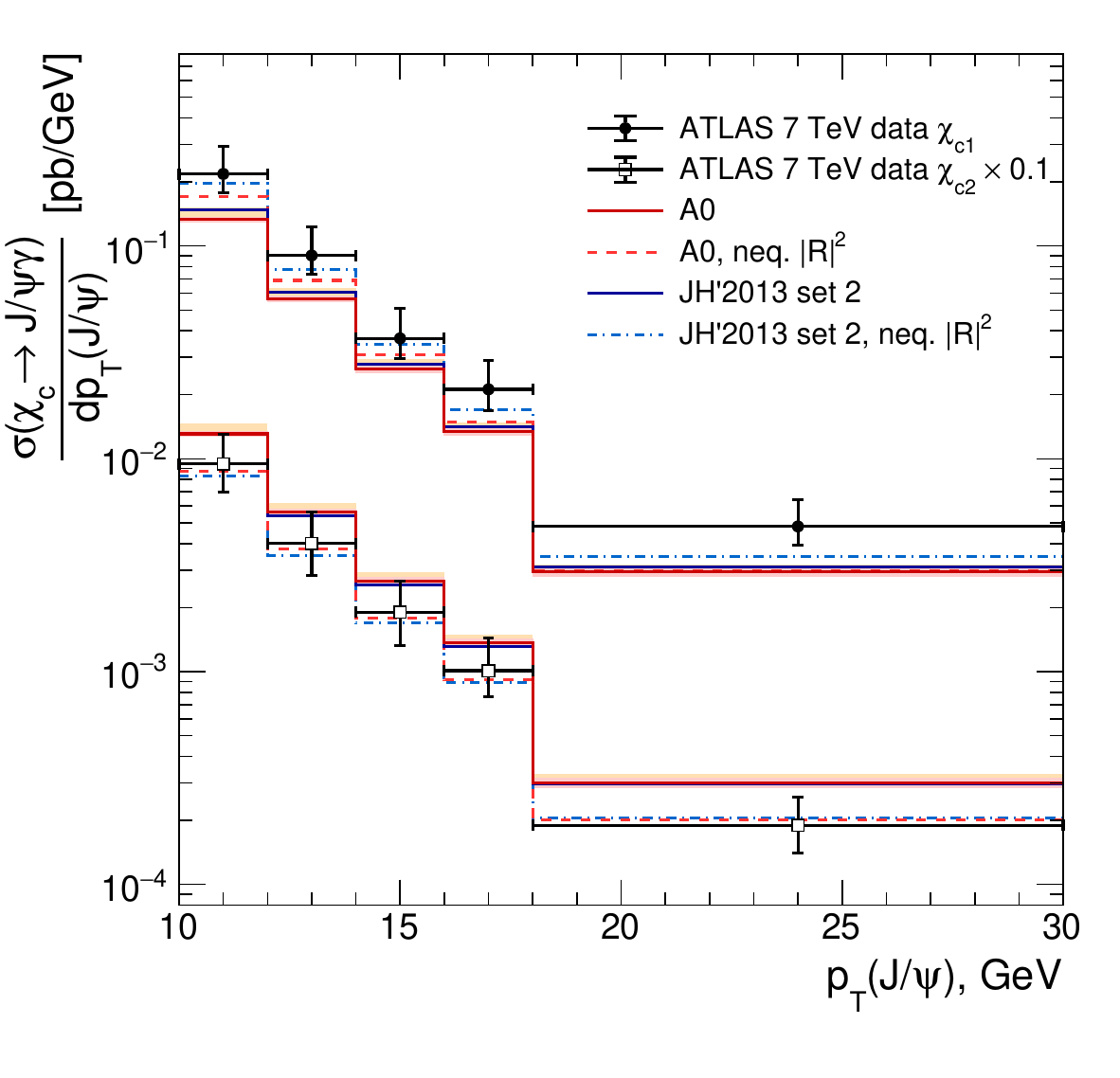}}\hfill
{\includegraphics[width=.49\textwidth]{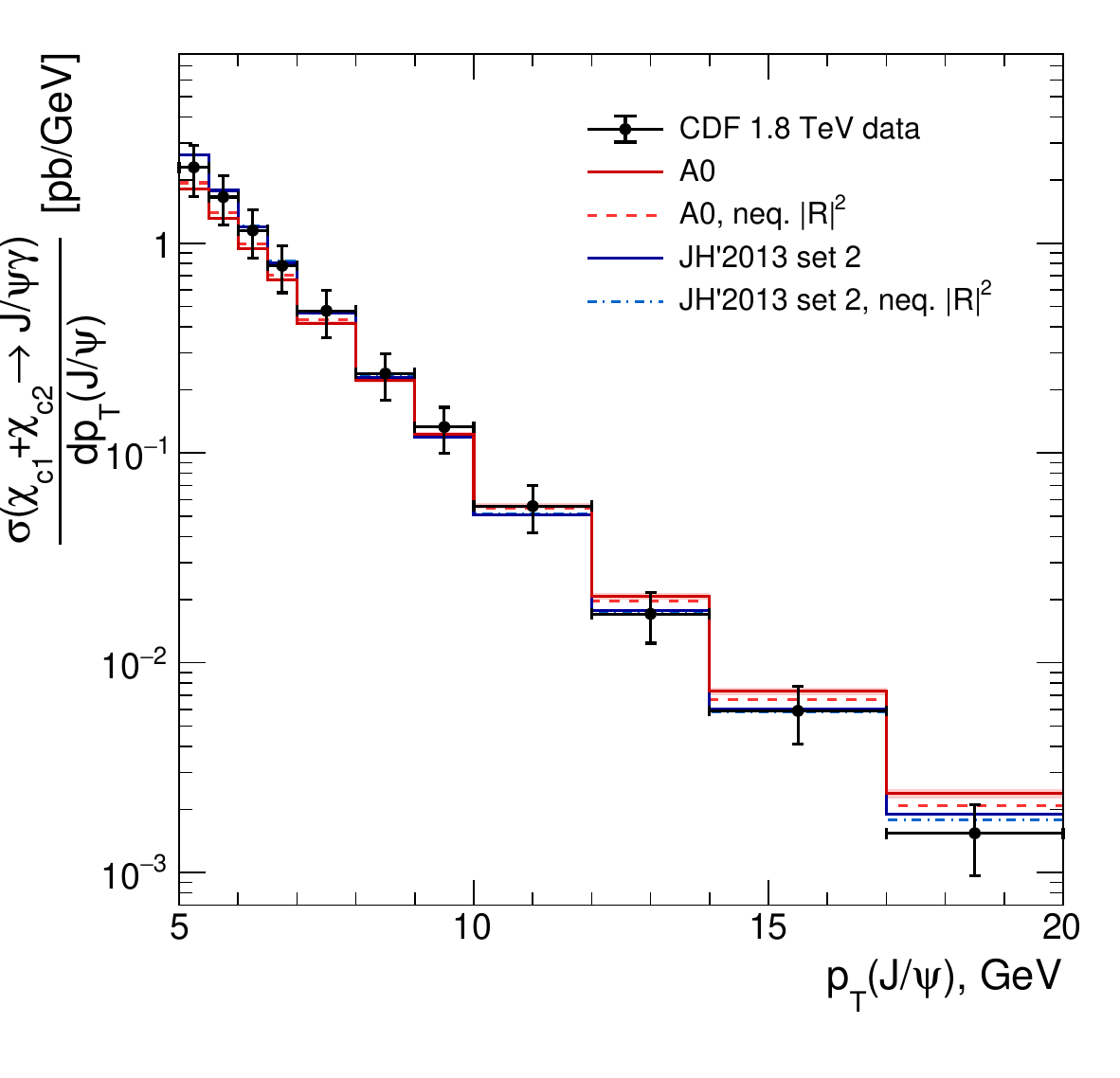}}\hfill
\caption{The differential cross sections for inclusive $\chi_{c1,2}$ production in $pp$ collisions at $\sqrt{s} = 7$ TeV  and $\sqrt{s} = 1.8$ TeV as functions of the $J/\psi$ transverse momentum at the forward rapidity regions. The ATLAS data are taken from \cite{dataset_chic_ATLAS7} and CDF \cite{dataset_chic_CDF}.}
\label{fig:ATLAS_CDF_chicJ}
\end{center}
\end{figure}

\begin{figure}
\begin{center}
{\includegraphics[width=.49\textwidth]{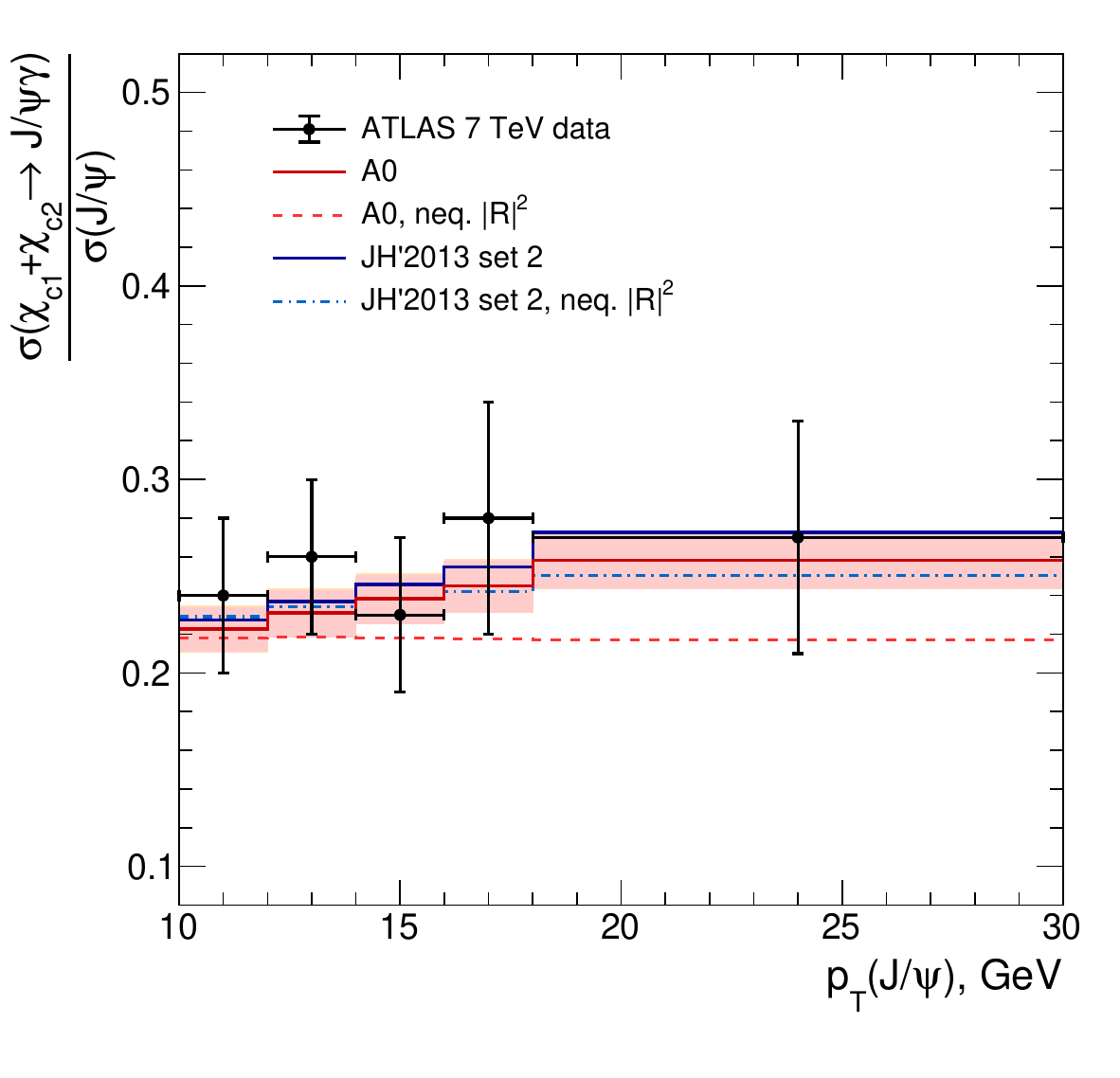}}\hfill
{\includegraphics[width=.49\textwidth]{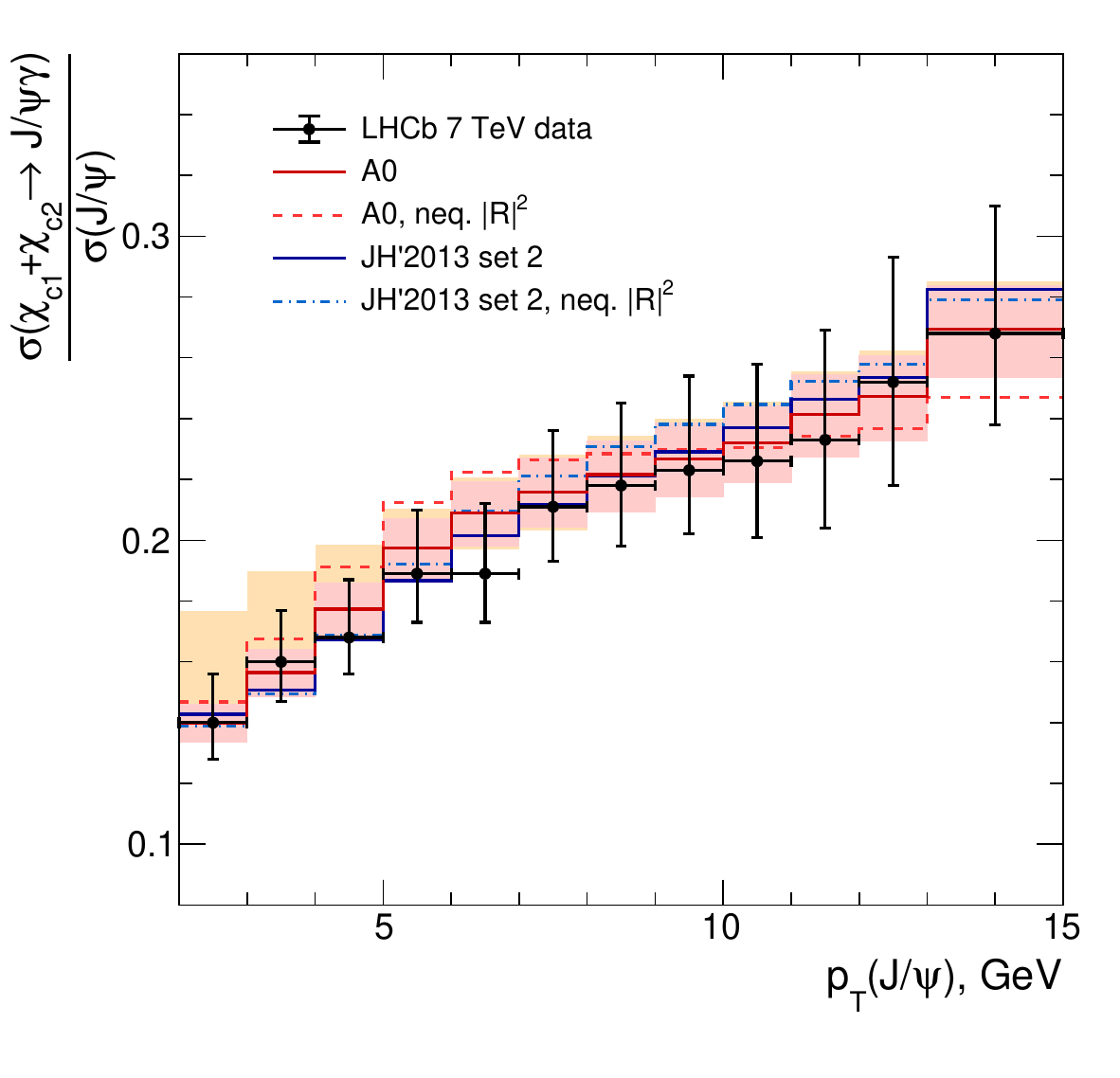}}\hfill
\caption{Ratios of differential cross sections  $\sigma(\chi_{c1}+\chi_{c2})/\sigma(J/\psi)$ in $pp$ collisions at $\sqrt{s} = 7$ TeV as functions of the $J/\psi$ transverse momentum at the central (left plot) and forward (right plot) rapidity regions. The ATLAS data are taken from \cite{dataset_chic_ATLAS7} and LHCb \cite{dataset_chic_LHCb2}.}
\label{fig:ATLAS_LHCb_chicJ_Jpsi}
\end{center}
\end{figure}

\begin{figure}
\begin{center}
{\includegraphics[width=.49\textwidth]{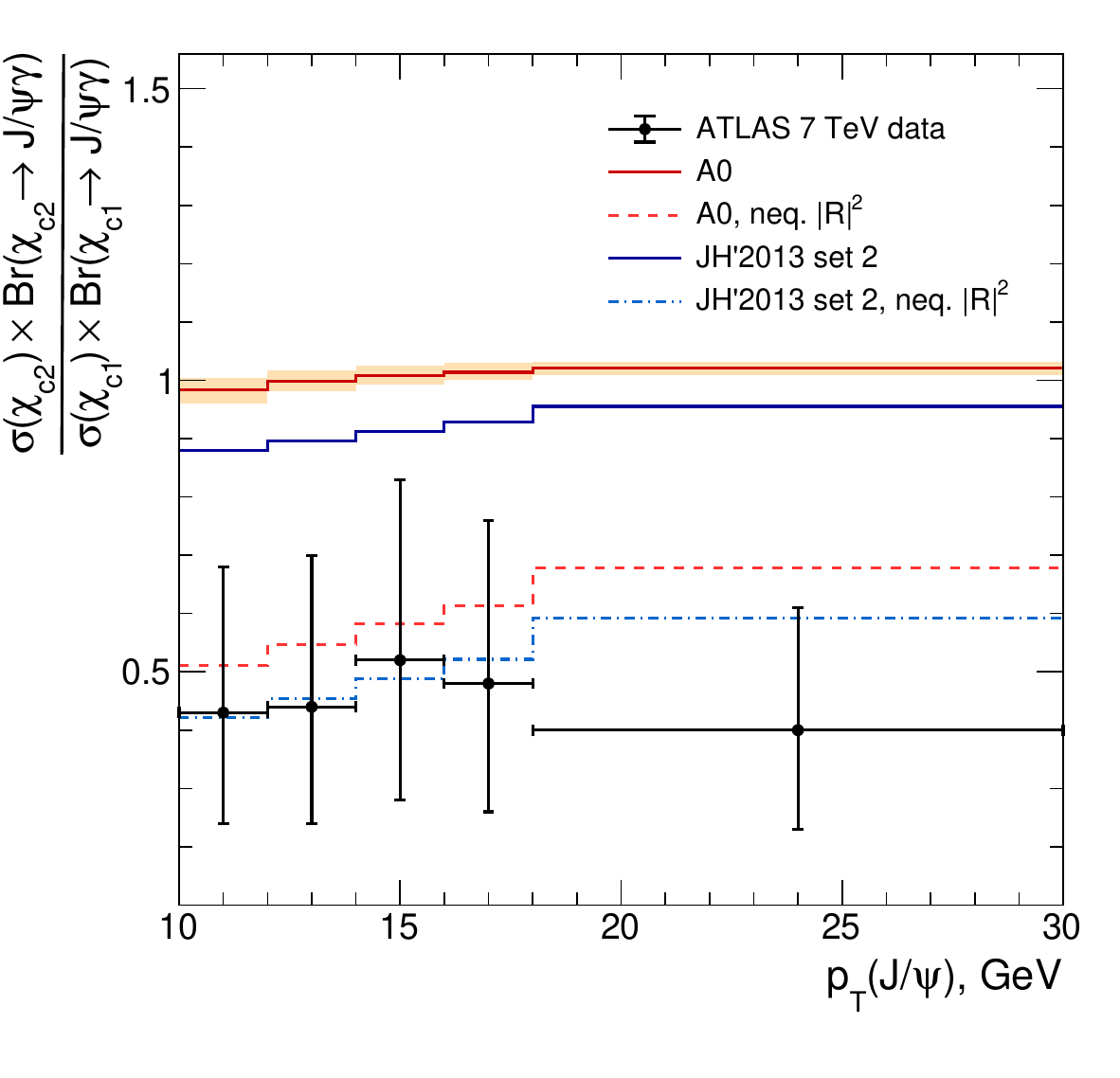}}\hfill
{\includegraphics[width=.49\textwidth]{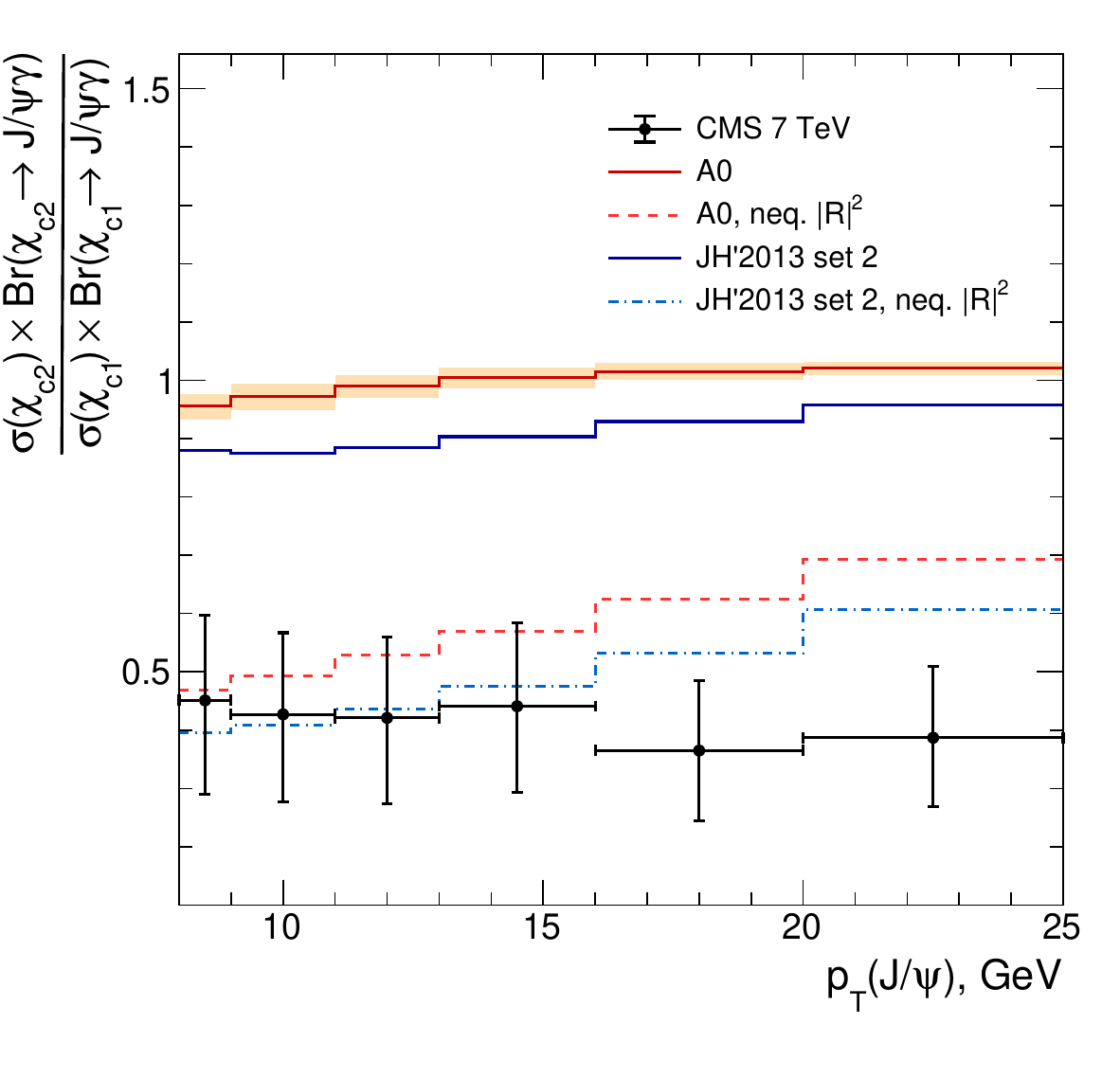}}\hfill
{\includegraphics[width=.49\textwidth]{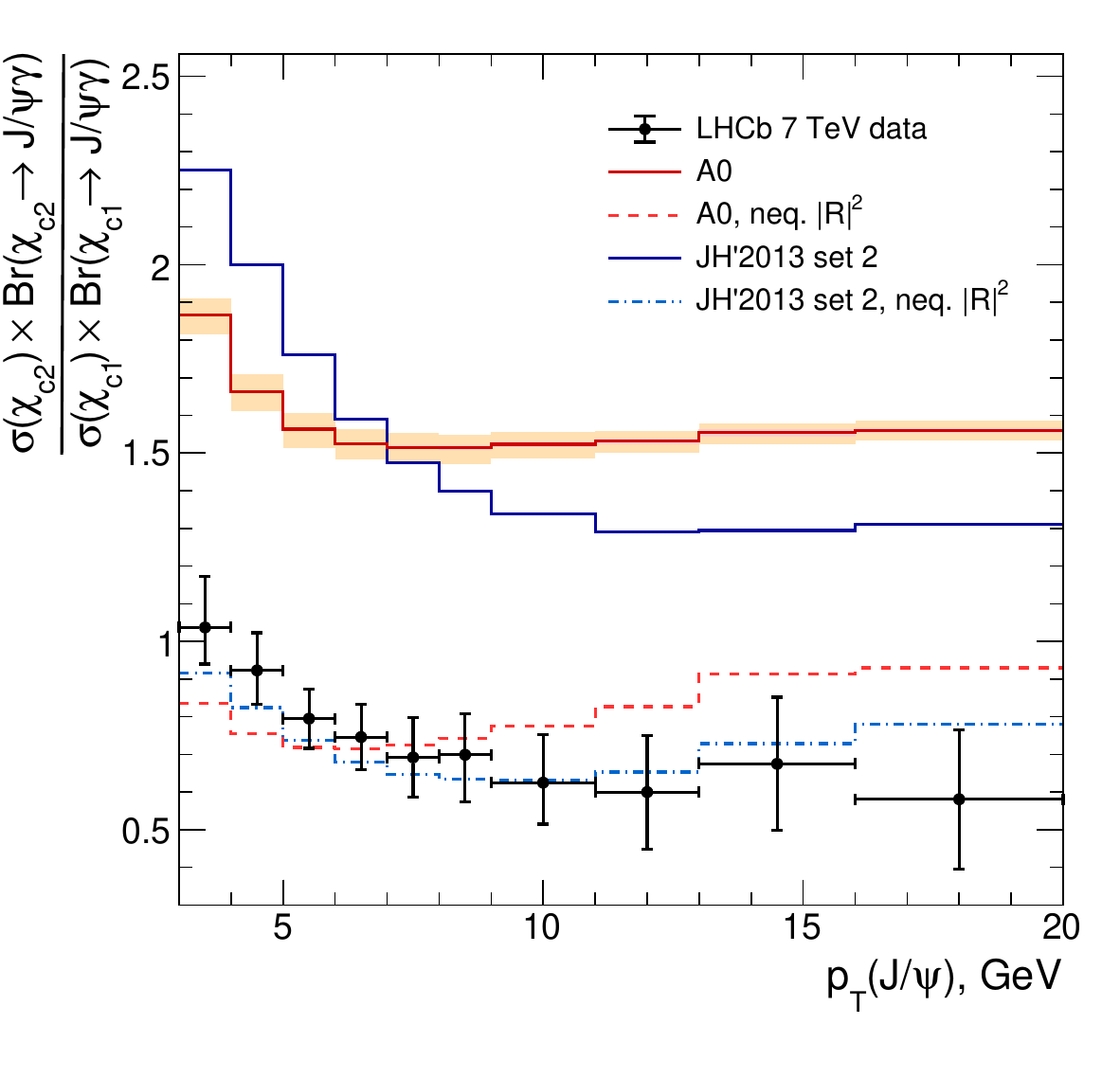}}\hfill
\caption{Ratios of differential cross sections  $\sigma(\chi_{c2})/\sigma(\chi_{c1})$ in $pp$ collisions at $\sqrt{s} = 7$ TeV as functions of the $J/\psi$ transverse momentum at the central (upper plots) and forward (lower plot) rapidity regions. The ATLAS data are taken from \cite{dataset_chic_ATLAS7}, CMS \cite{dataset_chic_CMS} and LHCb \cite{dataset_chic_LHCb}.}
\label{fig:ATLAS_LHCb_chicJ_Jpsi}
\end{center}
\end{figure}

\begin{table}[ht]
\centering
\small
\caption{${\chi}^{2}/{\rm n.d.f.}$ for available $J/\psi$ datasets at $\mathrm{LO}+\mathrm{NLO}^{\dag}$.}
\label{tab:Jpsi_chi2}
\vspace{0.2cm}
% Таблица растягивается на всю ширину страницы.
% Для первых трех колонок — фиксированная ширина (l, c, c), последние две (длинные заголовки) — умные центрированные X
\begin{tabularx}{\textwidth}{l c >{\centering\arraybackslash}X >{\centering\arraybackslash}X >{\centering\arraybackslash}X}
\toprule
Experiment & n.d.f. & A0 & JH'2013 set 2 & JH'2013 set 2 \newline ($p_{T} > 6$) \\
\midrule
ATLAS 5 TeV {\scriptsize \cite{dataset_ATLAS5}}   & 33      & 0.36 & 4.89 & 1.02 \\
ATLAS 7 TeV {\scriptsize \cite{dataset_ATLAS7_8}}  & 176     & 3.25 & 15.6 & 3.68 \\
ATLAS 8 TeV {\scriptsize \cite{dataset_ATLAS7_8}}  & 172     & 0.91 & 4.14 & 1.00 \\
ATLAS 13 TeV {\scriptsize \cite{dataset_ATLAS13}}  & 90      & 7.82 & 28.6 & 5.98 \\
CMS 5 TeV {\scriptsize \cite{dataset_CMS5}}       & 52(40)  & 2.38 & 24.8 & 6.01 \\
CMS 7 TeV {\scriptsize \cite{dataset_CMS7}}        & 120     & 1.59 & 5.39 & 0.41 \\
CMS 13 TeV {\scriptsize \cite{dataset_CMS13}}      & 84      & 0.66 & 7.10 & 1.91 \\
LHCb 5 TeV {\scriptsize \cite{dataset_LHCb5}}     & 50(20)  & 13.3 & 46.1 & 17.8 \\
LHCb 7 TeV {\scriptsize \cite{dataset_LHCb7}}   & 66(36)  & 4.56 & 13.3 & 5.05 \\
LHCb 8 TeV {\scriptsize \cite{dataset_LHCb8}}     & 70(40)  & 1.59 & 16.6 & 4.45 \\
LHCb 13 TeV {\scriptsize \cite{dataset_LHCb13}}  & 70(40)  & 12.1 & 25.8 & 5.66 \\

\midrule
\textbf{Summary}                     & \textbf{983(851)} & \textbf{3.81} & \textbf{15.1} & \textbf{3.27} \\
\bottomrule
\end{tabularx}
\end{table}
Similarly, Table ~\ref{tab:Jpsi_chi2} presents the $\chi ^{2}$ results for prompt $J/\psi$ production data, namely the differential cross sections $d\sigma(J/\psi)/dp_T$. These calculations are also performed using the LDMEs listed in Table 2. The results for the scenario with independent $|\mathcal{R}^{\prime\chi_{cJ}}(0)|^2$ values are not presented separately because both scenarios lead to nearly identical outcomes. This is not surprising since the first stage of the fitting procedure rigidly constrained the linear combination of the LDMEs for $\chi _{cJ}$ and $J/\psi$ mesons $\eqref{eq:linear_comb_Jpsi_chic1}-\eqref{eq:linear_comb_Jpsi_chic2}$, allowing the parameters to vary freely but fixing the total contribution from direct $J/\psi$ and the feed-down contributions from $\chi _{cJ}$ decays. Similar to the situation with the $\psi^{\prime}$ analysis, the largest contribution to the total $\chi ^{2}$ comes from the LHCb data. In addition, a discrepancy also remains between the results from the ATLAS and CMS Collaborations taken at $\sqrt s = 13$~TeV.

The results of our calculations for $J/\psi$ transverse momentum spectra are presented in Fig.~\ref{fig:ATLAS_7TeV} -- \ref{fig:LHCb_8_13TeV} for all datasets in the form of ``Theory/Data'' ratios. Both
the A0 and JH'2013 set 2 (with the $p_T > 6$ GeV cutoff) gluon densities demonstrate good agreement with all available experimental data. Separately, we focus on the comparison with the LHCb data shown in Figs.~\ref{fig:LHCb_5_7TeV} and~\ref{fig:LHCb_8_13TeV}. One can observe a visible trend from an overestimation of the data at an energy of 5~TeV toward an underestimation at 13~TeV, which might indicate an imperfect energy scaling in the forward rapidity region for the gluon distribution functions employed in this work.
Nevertheless, our predictions are rather close to the measurements
within the estimated theoretical and experimental uncertainties.

\begin{figure}
\begin{center}
{\includegraphics[width=.49\textwidth]{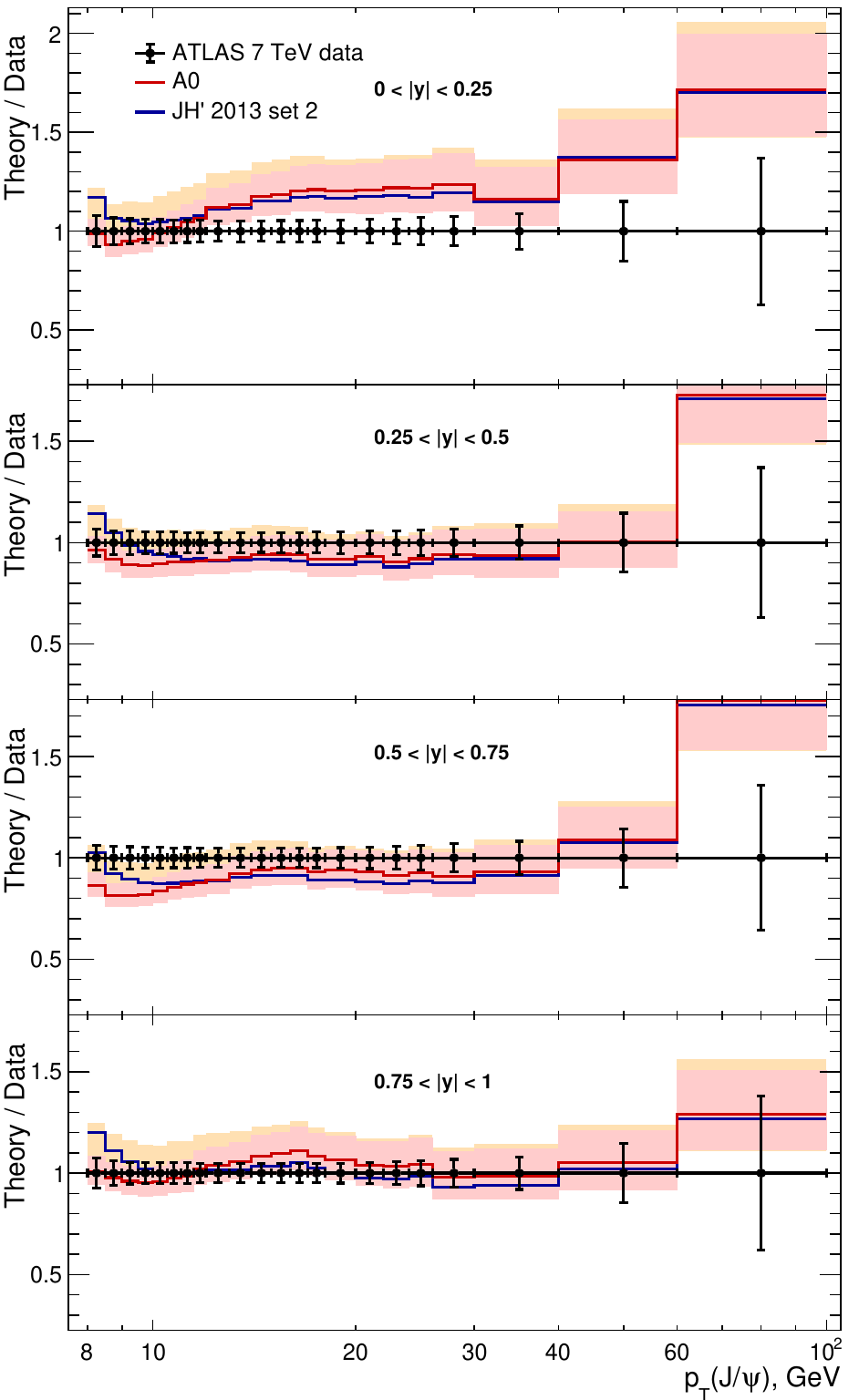}}\hfill
{\includegraphics[width=.49\textwidth]{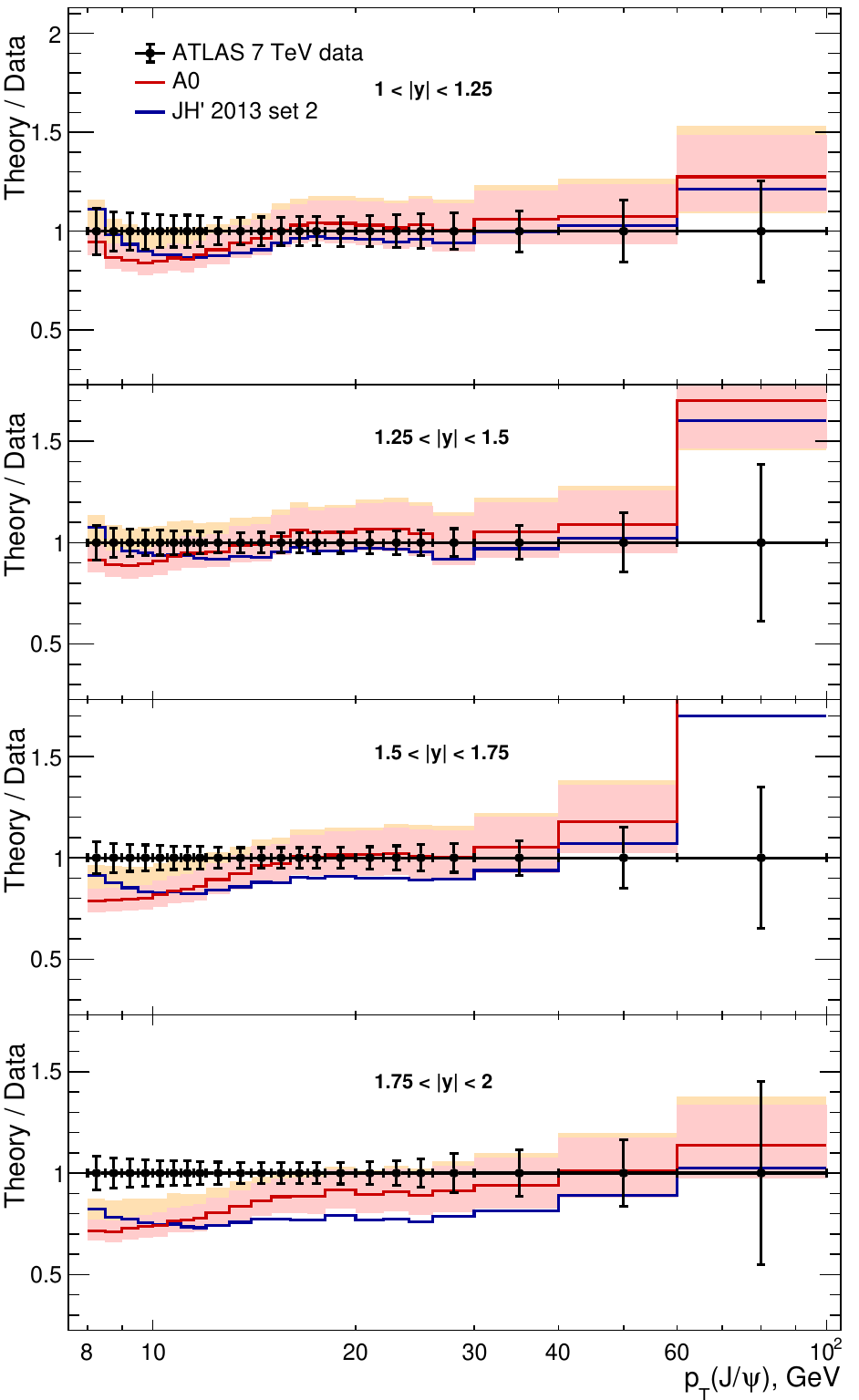}}\hfill
\caption{Ratios of differential cross sections for inclusive $J/\psi$ production in $pp$ collisions at $\sqrt{s} = 7$ TeV as functions of the $J/\psi$ transverse momentum at the central rapidity regions. The ATLAS data are taken from \cite{dataset_ATLAS7_8}.}
\label{fig:ATLAS_7TeV}
 \end{center}
\end{figure}

\begin{figure}
\begin{center}
{\includegraphics[width=.49\textwidth]{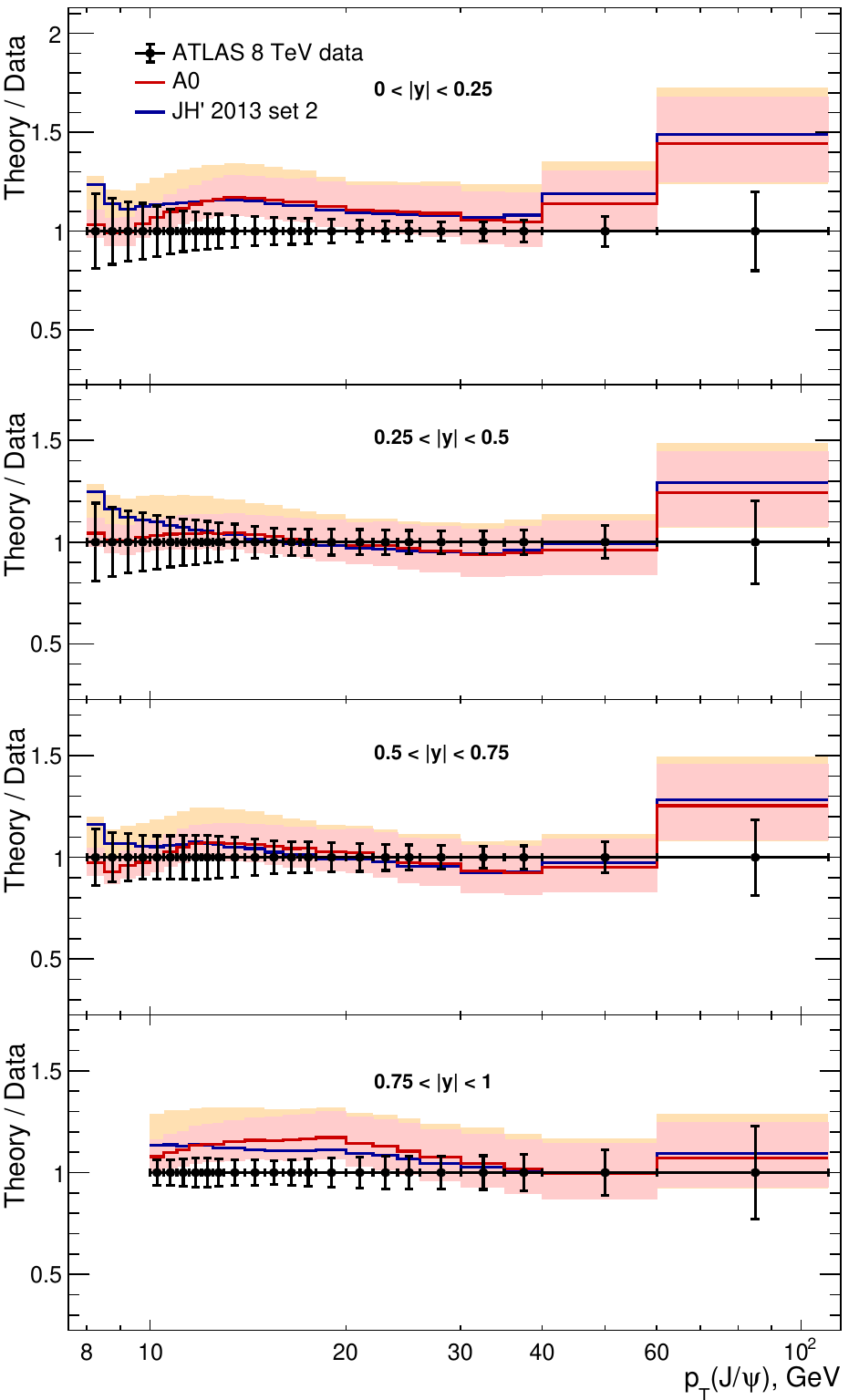}}\hfill
{\includegraphics[width=.49\textwidth]{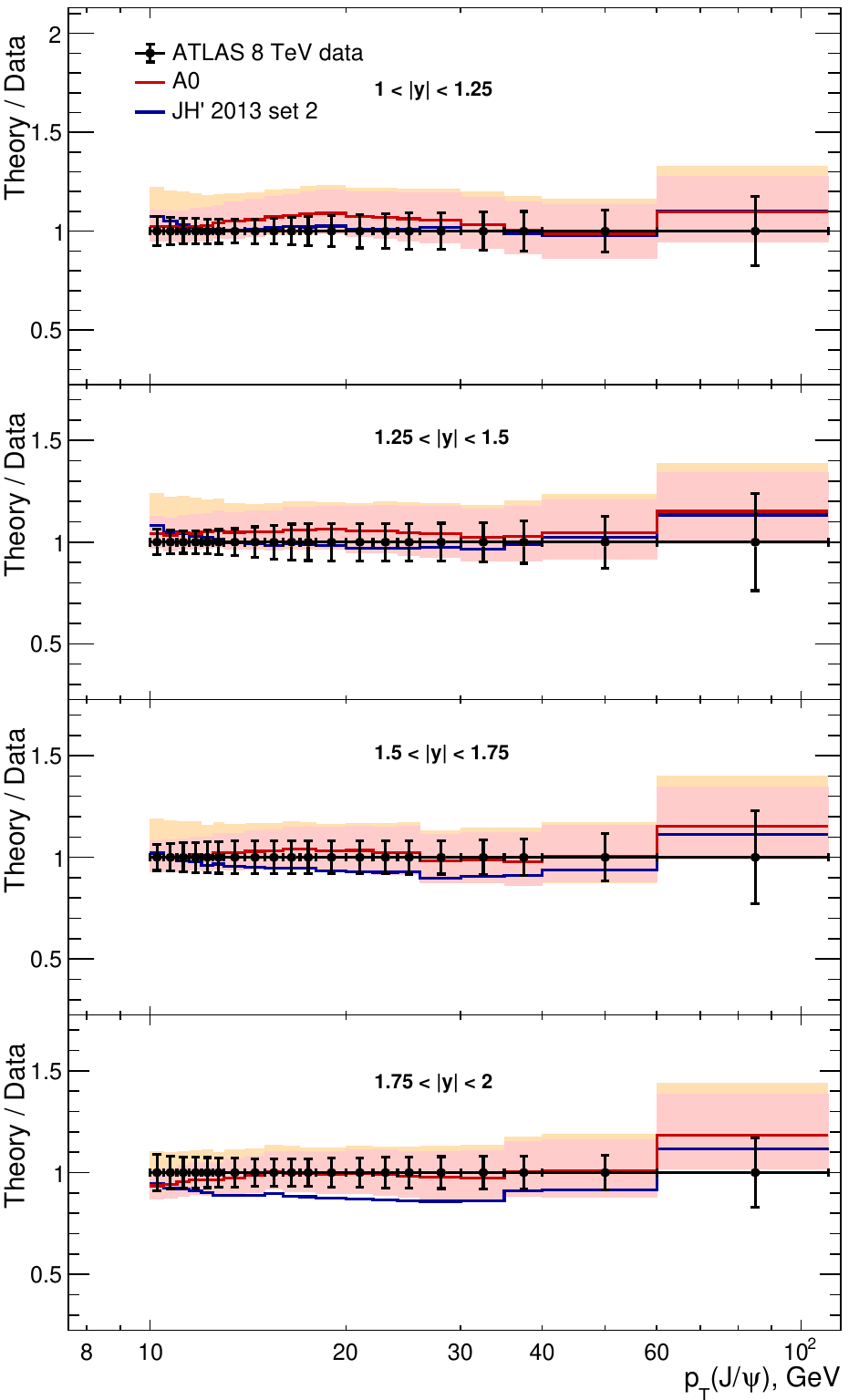}}\hfill
\caption{Ratios of differential cross sections for inclusive $J/\psi$ production in $pp$ collisions at $\sqrt{s} = 8$ TeV as functions of the $J/\psi$
transverse momentum at the central rapidity regions. The ATLAS data are taken from \cite{dataset_ATLAS7_8}.}
\label{fig:ATLAS_8TeV}
\end{center}
\end{figure}

\begin{figure}
\begin{center}
{\includegraphics[width=.49\textwidth]{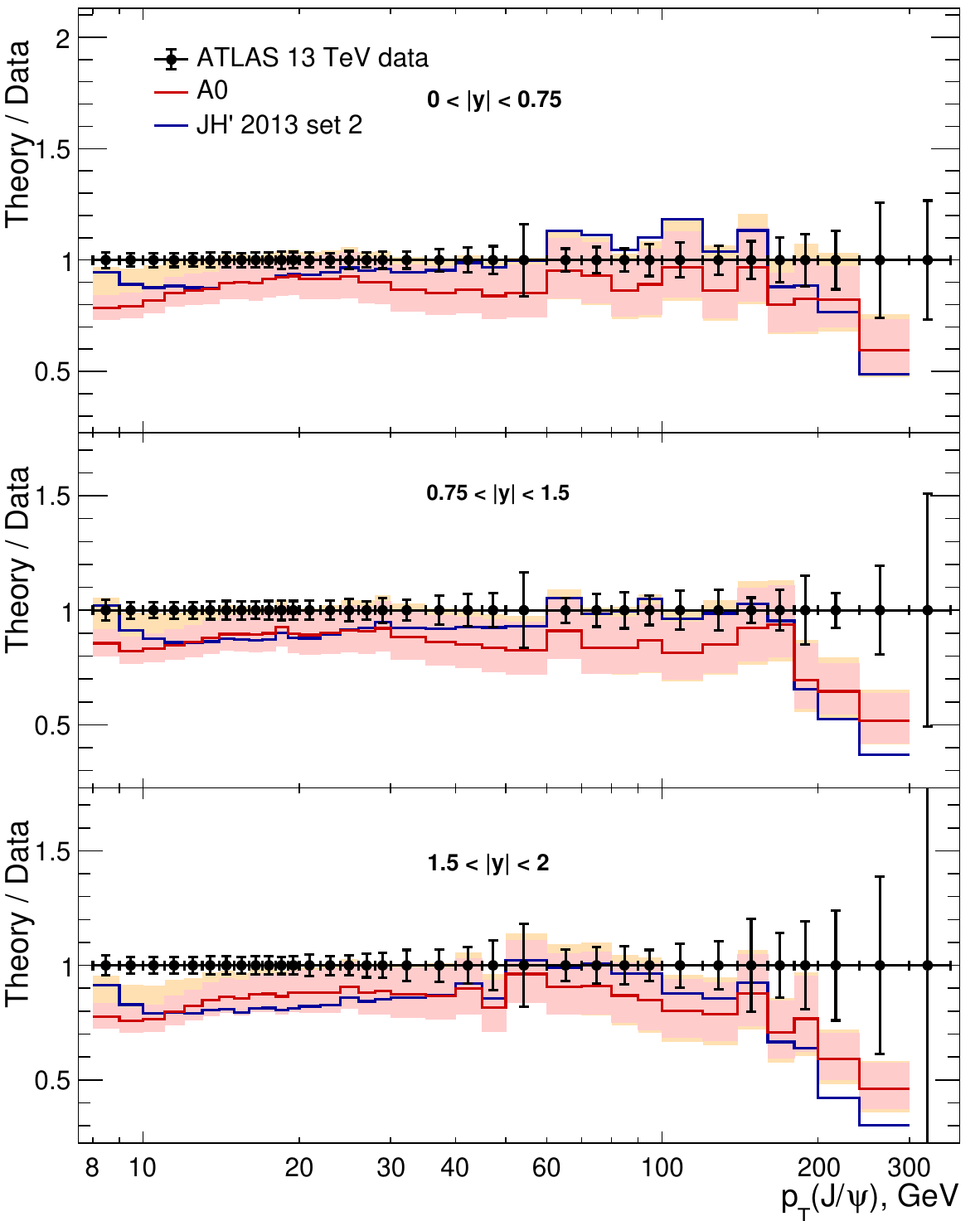}}\hfill
{\includegraphics[width=.49\textwidth]{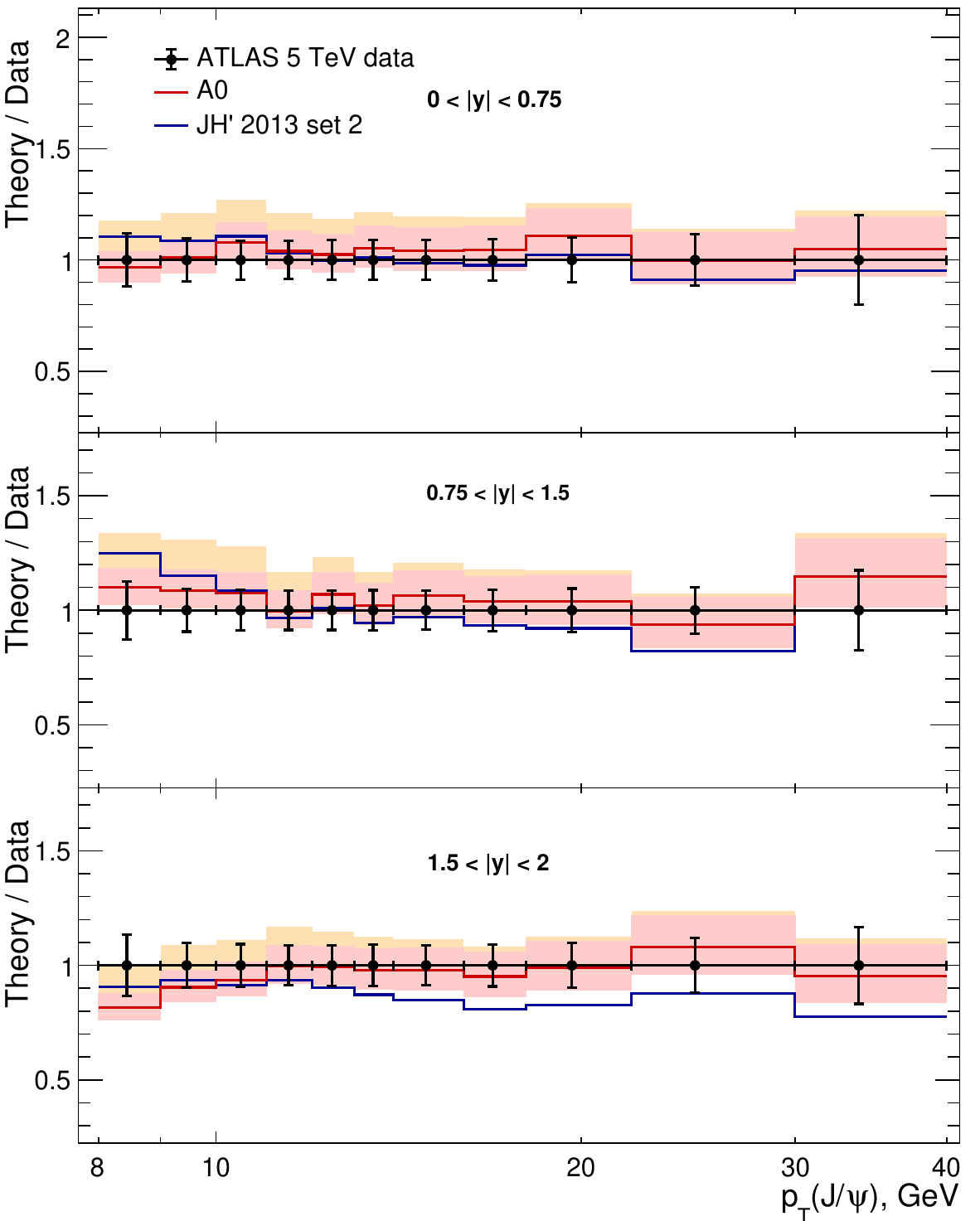}}\hfill
\caption{Ratios of differential cross sections for inclusive $J\psi$ production in $pp$ collisions at $\sqrt{s} = 13$ TeV  and $\sqrt{s} = 5$ TeV as functions of the $J/\psi$ transverse momentum at the central rapidity regions. The ATLAS data are taken from \cite{dataset_ATLAS13} and \cite{dataset_ATLAS5}.}
\label{fig:ATLAS_5_13TeV}
\end{center}
\end{figure}

\begin{figure}
\begin{center}
{\includegraphics[width=.49\textwidth]{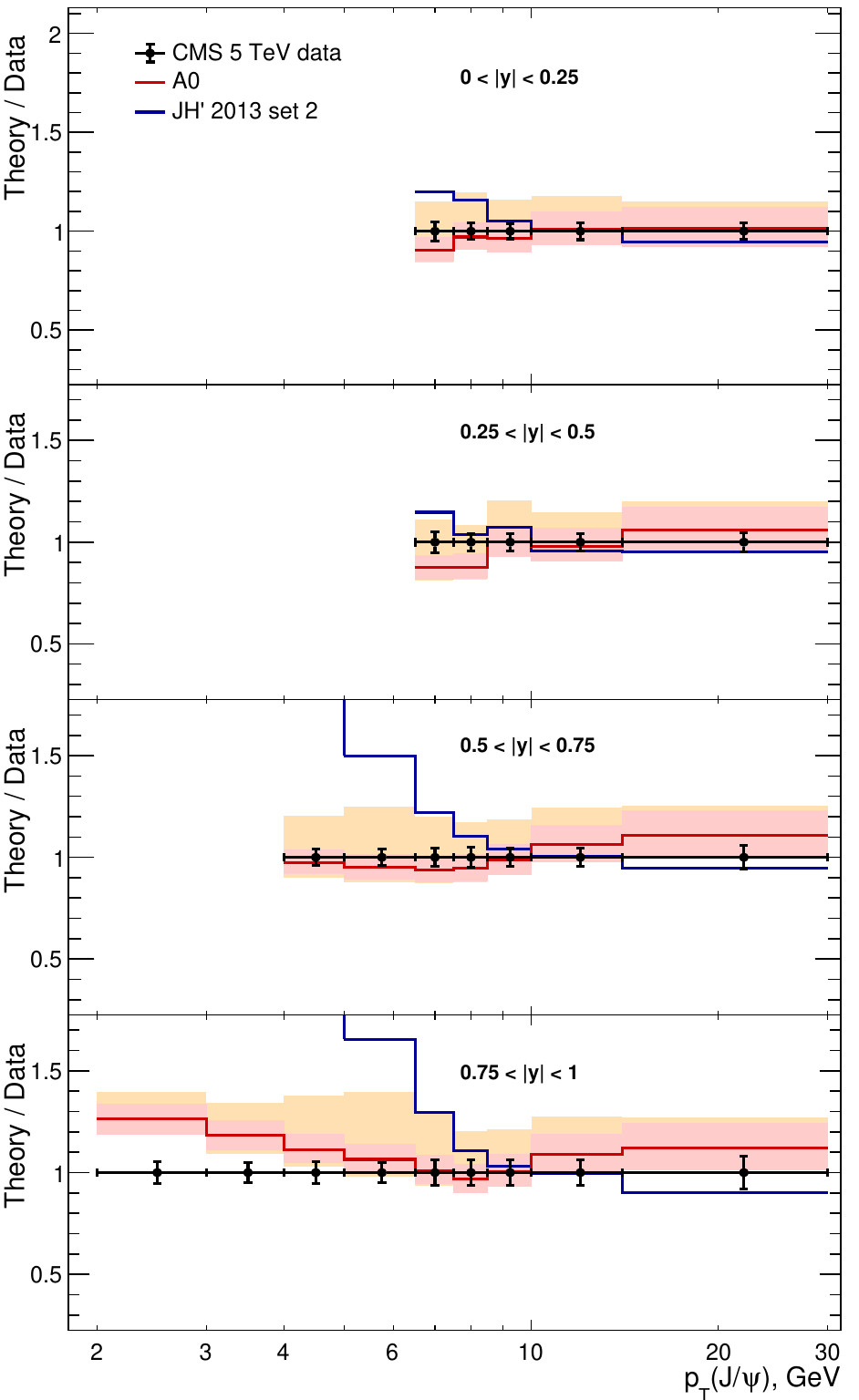}}\hfill
{\includegraphics[width=.49\textwidth]{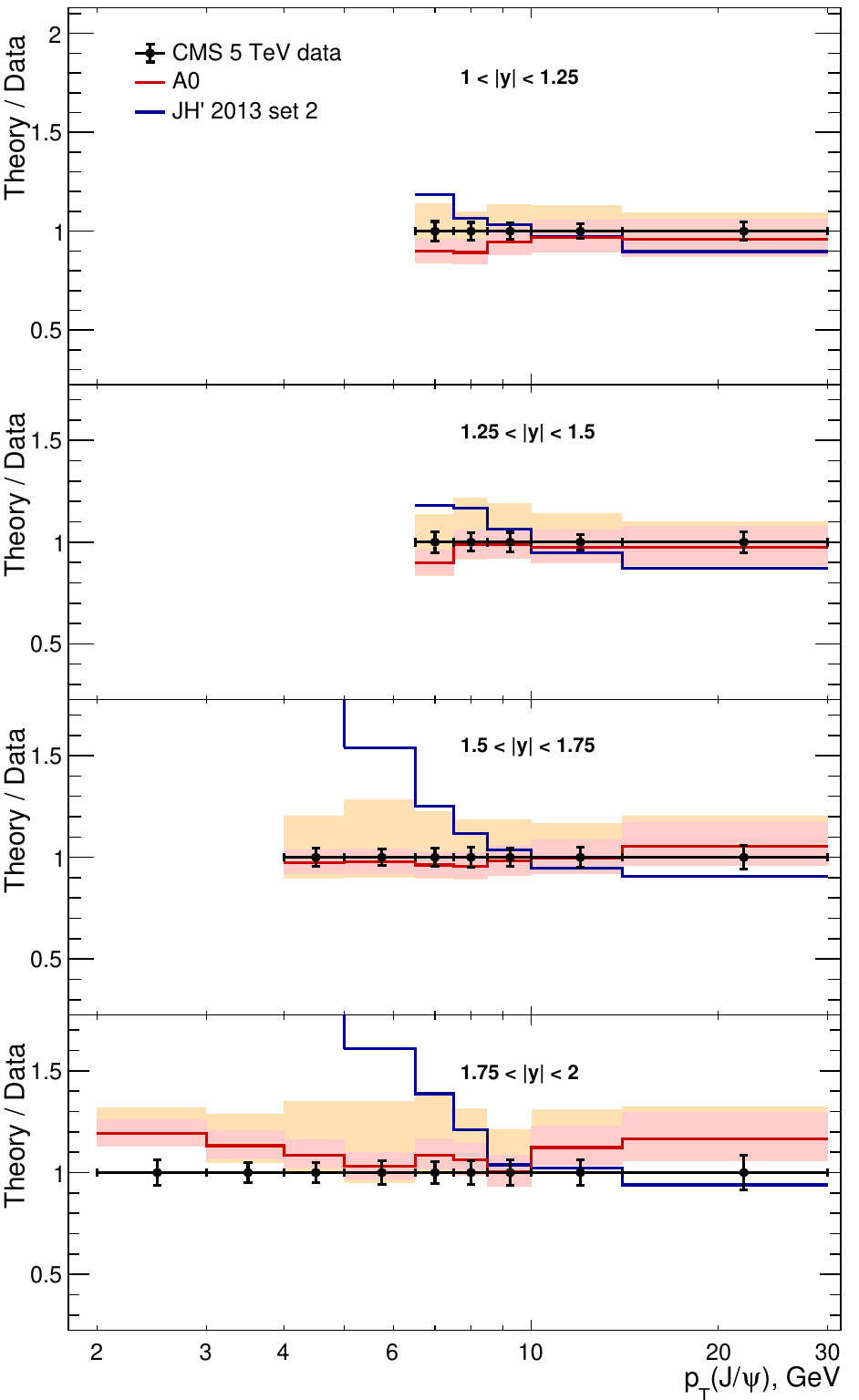}}\hfill
\caption{Ratios of differential cross sections for inclusive $J/\psi$ production in $pp$ collisions at $\sqrt{s} = 5$ TeV at the central rapidity regions as functions of the $J/\psi$ transverse momentum. The CMS data are taken from \cite{dataset_CMS5}.}
\label{fig:CMS_5}
\end{center}
\end{figure}

\begin{figure}
\begin{center}
{\includegraphics[width=.49\textwidth]{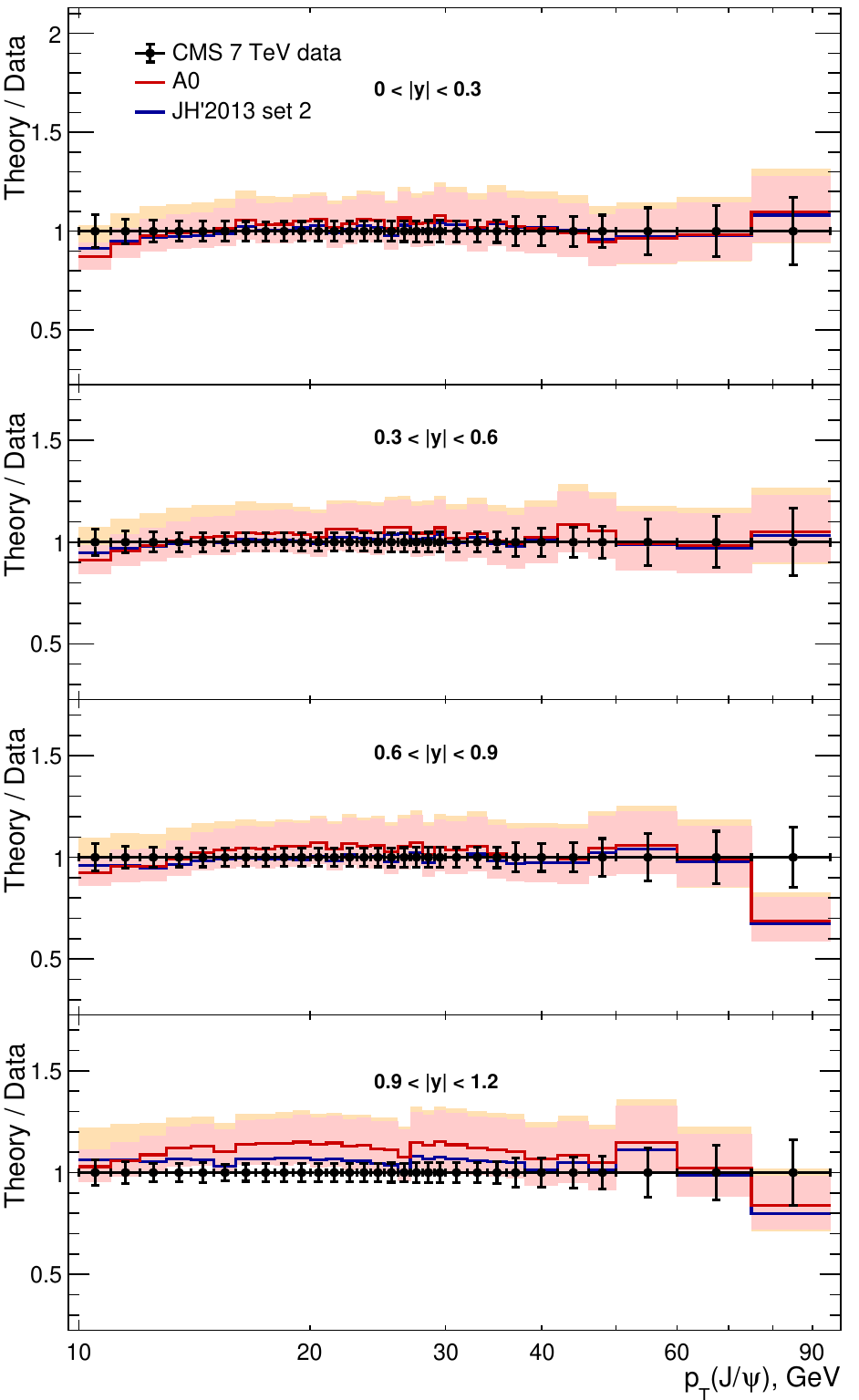}}\hfill
{\includegraphics[width=.49\textwidth]{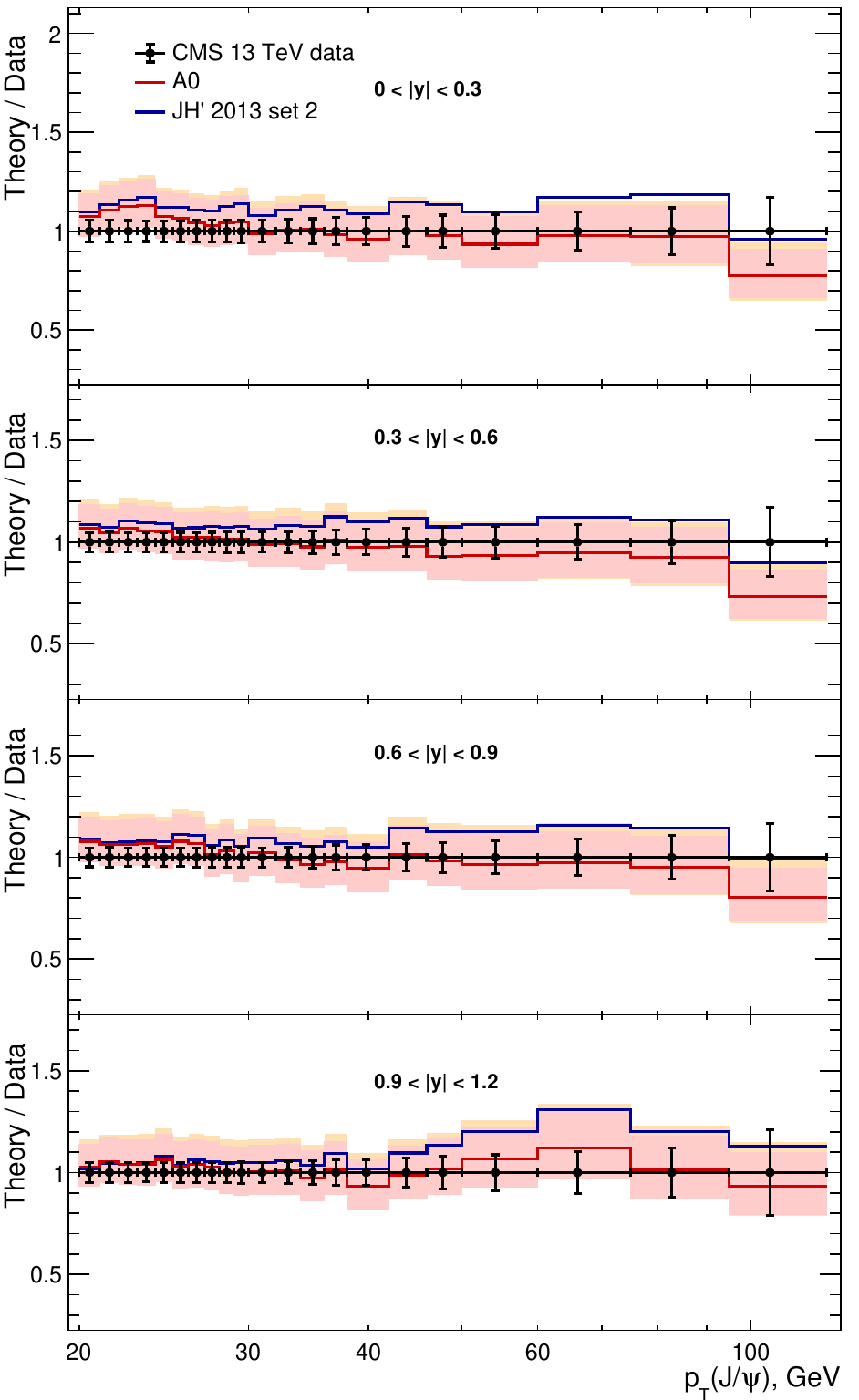}}\hfill
\caption{Ratios of differential cross sections for inclusive $J\psi$ production in $pp$ collisions at $\sqrt{s} = 7$ TeV  and $\sqrt{s} = 13$ TeV as functions of the $J/\psi$ transverse momentum at the central rapidity regions. The CMS data are taken from \cite{dataset_CMS7} and \cite{dataset_CMS13}.}
\label{fig:CMS_7_13TeV}
\end{center}
\end{figure}

\begin{figure}
\begin{center}
{\includegraphics[width=.49\textwidth]{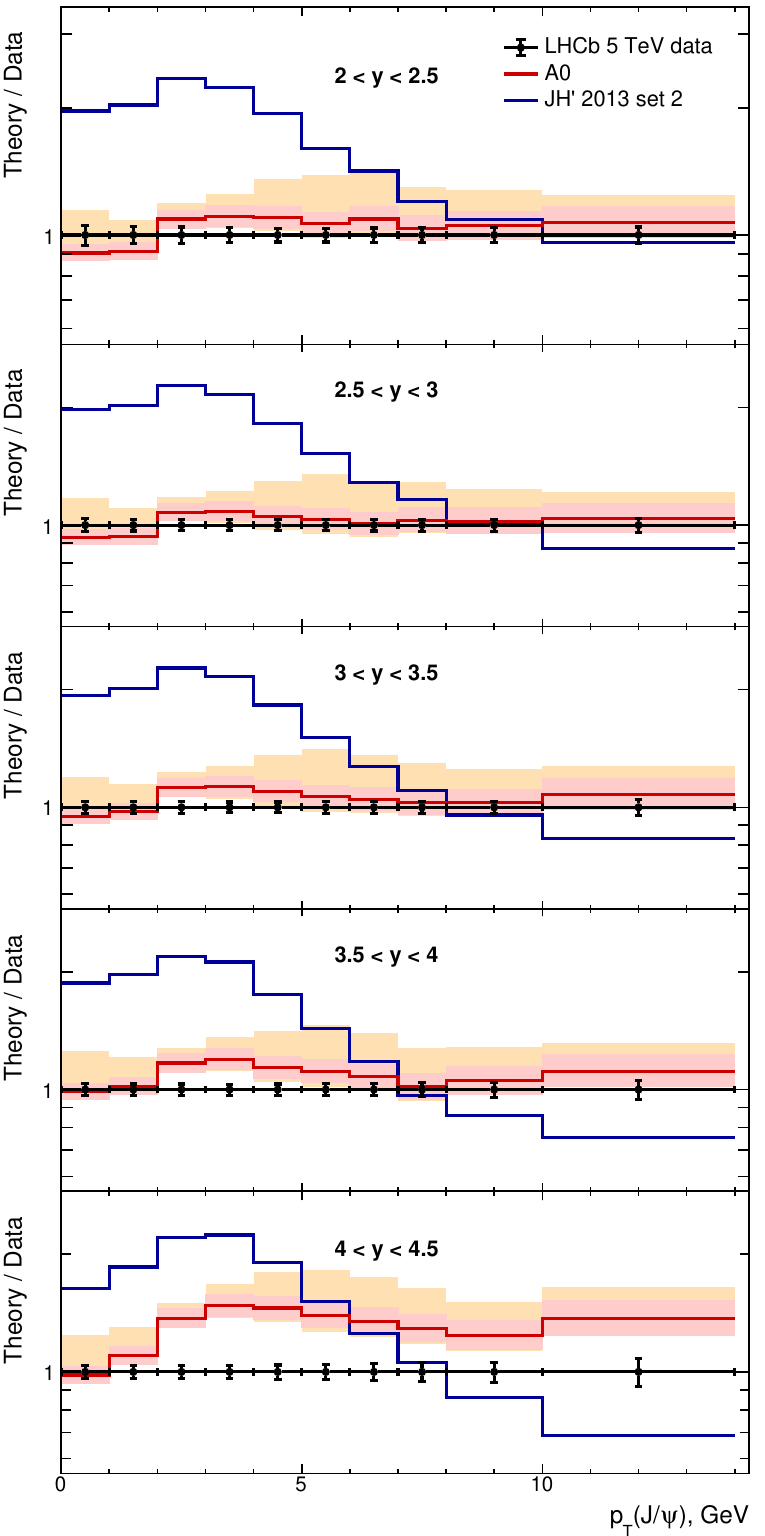}}\hfill
{\includegraphics[width=.49\textwidth]{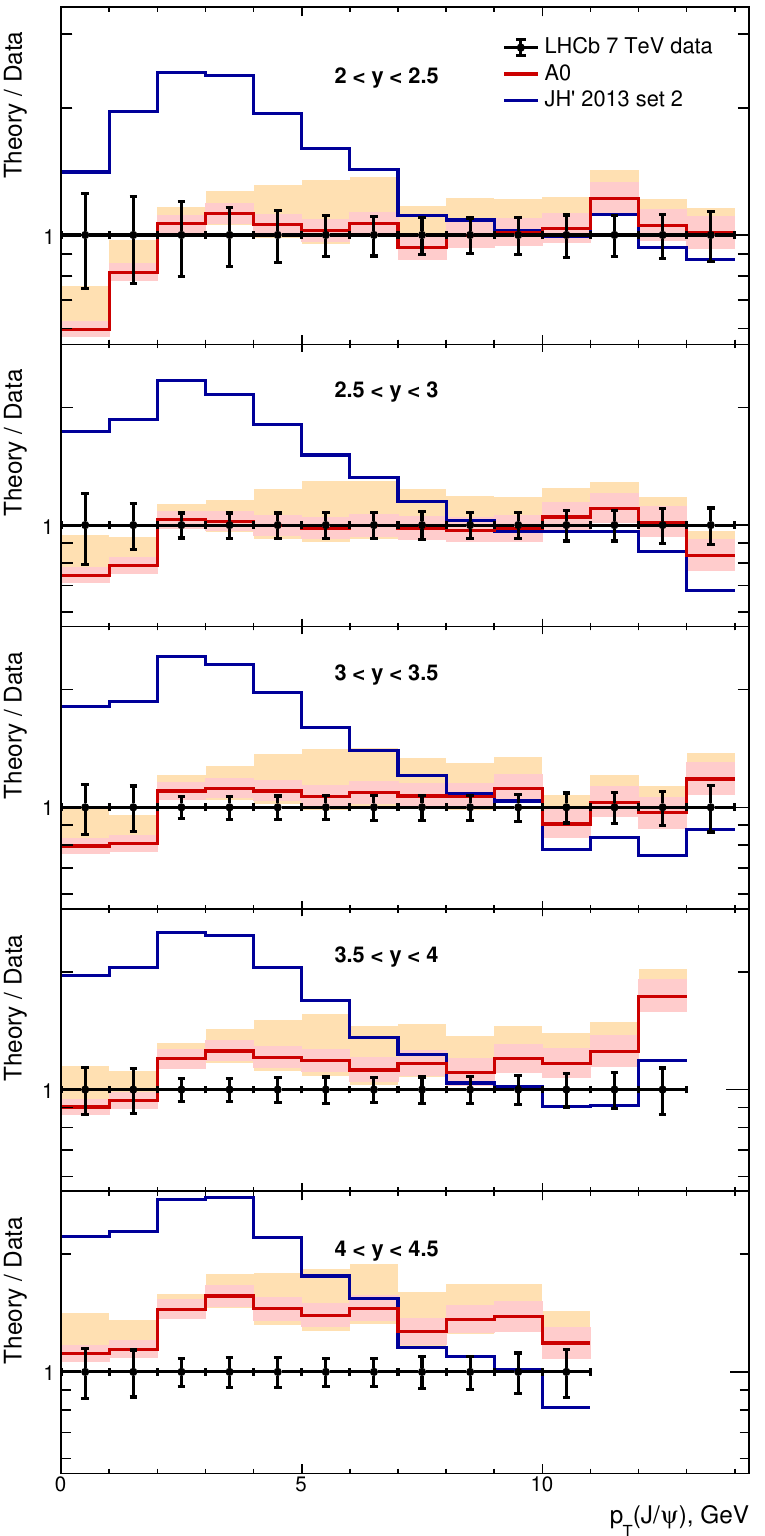}}\hfill
\caption{Ratios of differential cross sections for inclusive $J\psi$ production in $pp$ collisions at $\sqrt{s} = 5$ TeV  and $\sqrt{s} = 7$ TeV as functions of the $J/\psi$ transverse momentum at the forward rapidity regions. The LHCb data are taken from \cite{dataset_LHCb5} and \cite{dataset_LHCb7}.}
\label{fig:LHCb_5_7TeV}
\end{center}
\end{figure}

\begin{figure}
\begin{center}
{\includegraphics[width=.49\textwidth]{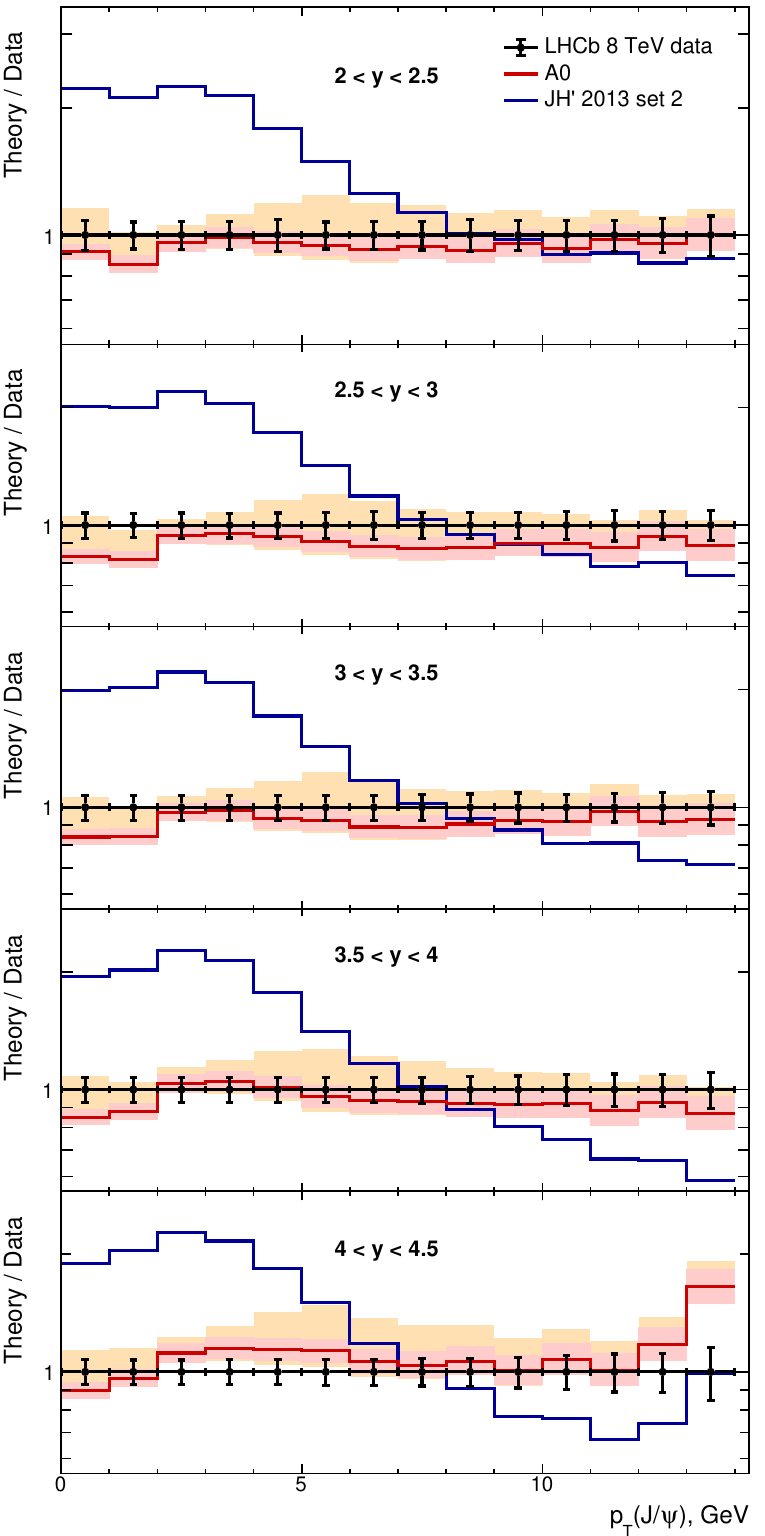}}\hfill
{\includegraphics[width=.49\textwidth]{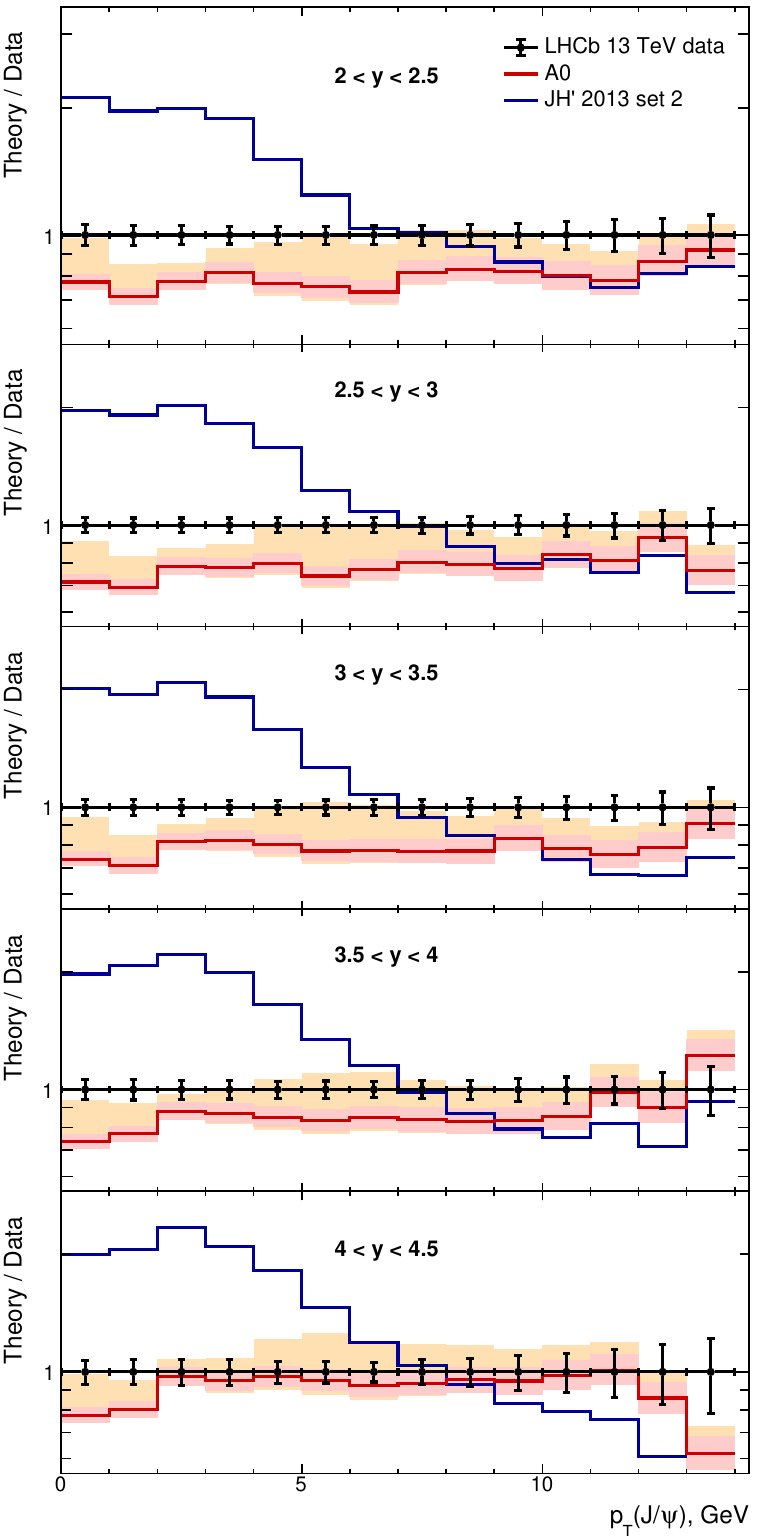}}\hfill
\caption{Ratios of differential cross sections for inclusive $J\psi$ production in $pp$ collisions at $\sqrt{s} = 8$ TeV  and $\sqrt{s} = 13$ TeV as functions of the $J/\psi$ transverse momentum at the forward rapidity regions. The LHCb data are taken from \cite{dataset_LHCb8} and \cite{dataset_LHCb13}.}
\label{fig:LHCb_8_13TeV}
\end{center}
\end{figure}

Additionally, to validate our predictions, we turn to the $\sigma(\psi^{\prime})/\sigma(J/\psi)$ relative production rates. These observables were not included in any fitting procedure, as they are not entirely independent, but rather derived from the direct measurements of $\psi^{\prime}$ and $J/\psi$ production. Although a large volume of data is available for analysis, for the sake of brevity, we present a comparison with only two datasets, namely, ATLAS~\cite{dataset_ATLAS13} and LHCb~\cite{dataset_LHCb_psi2S_jpsi} at $\sqrt{s}=13$ TeV, which cover the widest kinematic regions. As seen from the comparison in Fig.~\ref{fig:ATLAS_LHCb_psi_jpsi}, the measured ratios are well reproduced in both the central and forward rapidity regions. Notably, a significant reduction in the theoretical uncertainty bands is observed for these predictions, since such uncertainties systematically cancel out due to correlations. Consequently, it can be concluded that the proposed approach, {which incorporates tree-level NLO corrections, is able to provide a consistent and simultaneous description of all the available data for
transverse momenta distributions of unpolarized prompt $\psi^{\prime}$, $\chi _{cJ}$ and $J/\psi$ production in $pp$ collisions at high energies.

\begin{figure}
\begin{center}
{\includegraphics[width=.49\textwidth]{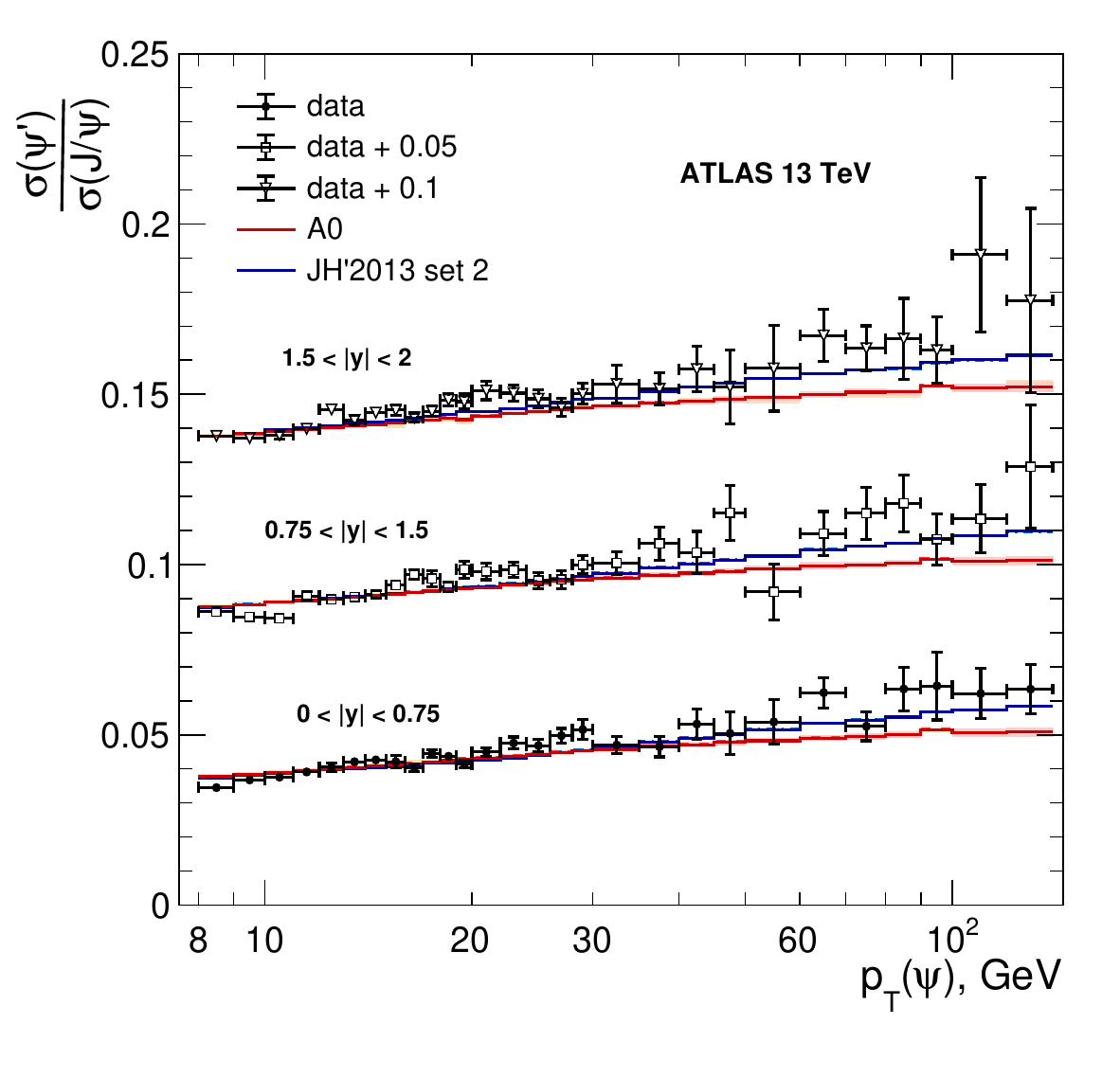}}\hfill
{\includegraphics[width=.49\textwidth]{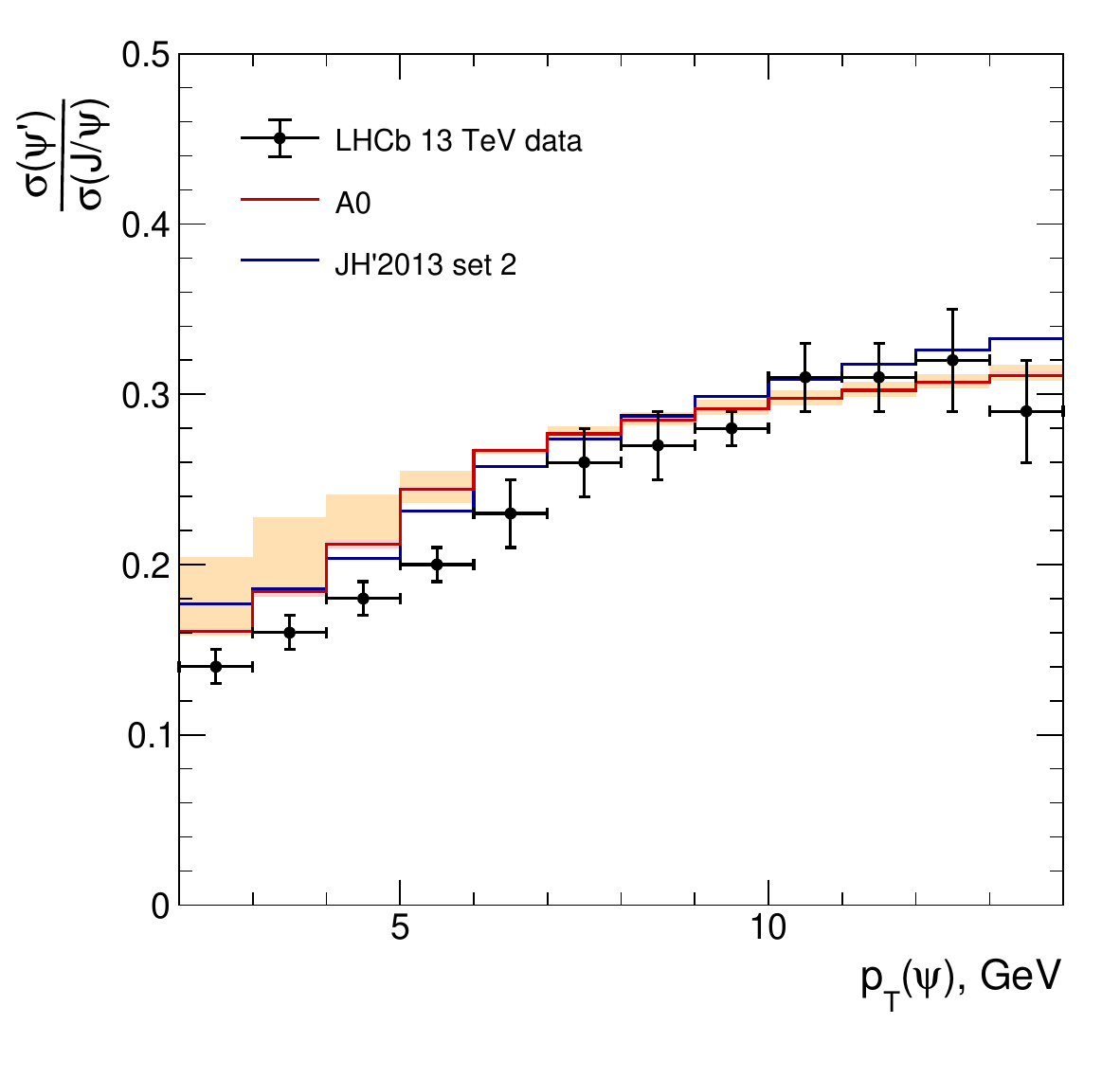}}\hfill
\caption{Ratios of differential cross sections  $\sigma(\psi^{\prime})/\sigma(J/\psi)$ in $pp$ collisions at $\sqrt{s} = 13$ TeV as functions of the $\psi$ transverse momentum at the central (left plot) and forward (right plot) rapidity regions. The ATLAS data are taken from \cite{dataset_ATLAS13} and LHCb \cite{dataset_LHCb_psi2S_jpsi}.}
\label{fig:ATLAS_LHCb_psi_jpsi}
\end{center}
\end{figure}

Finally, we turn to the $\eta_{c}$ hadroproduction at the LHC. As mentioned, the corresponding CO LDMEs could be determined from the ones for $J/\psi$ mesons via HQSS relations. Since the fitting procedure for $J/\psi$ required the introduction of a linear combination (\ref{eq:linear_comb_jpsi}) for the $\langle\mathcal{O}^{J/\psi}(^3P^{[8]}_0)\rangle$ and $\langle\mathcal{O}^{J/\psi}(^1S^{[8]}_0)\rangle$, we cannot explicitly calculate $\langle\mathcal{O}^{\eta_{c}}(^3S^{[8]}_1)\rangle$ and $\langle\mathcal{O}^{\eta_{c}}(^1P^{[8]}_1)\rangle$, but can only determine the allowed range for them.
Our calculations show that $^1S^{[1]}_0$ contribution alone significantly overestimates the experimental cross sections measured only in the forward rapidity region by LHCb Collaboration~\cite{dataset_etac_LHCb7_8, dataset_etac1_LHCb13, dataset_etac2_LHCb13}. Among the CO states, only the $^3S^{[8]}_1$ term can provide a non-negligible contribution\footnote{These results were also observed by other groups; see, for example,\cite{HQSSPuzzle-2}.}; thus, one could, for instance, set $\langle\mathcal{O}^{\eta_{c}}(^3S^{[8]}_1)\rangle = 0$ and $\langle\mathcal{O}^{\eta_{c}}(^1P^{[8]}_1)\rangle = 3\langle\mathcal{O}^{J/\psi}_{l.c.}\rangle$ to achieve better agreement with data.

Figure \ref{fig:LHCb_etac} presents a comparison of our calculations  with LHCb experimental
data collected at $\sqrt s = 7$, $8$ and $13$~TeV~\cite{dataset_etac_LHCb7_8, dataset_etac1_LHCb13, dataset_etac2_LHCb13}.
The theoretical uncertainties shown as shaded areas reflect only varying
$\langle\mathcal{O}^{\eta_{c}}(^1S^{[1]}_0)\rangle$. 
The results obtained with $\langle\mathcal{O}^{\eta_{c}}(^3S^{[8]}_1)\rangle = \langle\mathcal{O}^{J/\psi}_{l.c.}\rangle/r$ and $\langle\mathcal{O}^{\eta_{c}}(^1P^{[8]}_1)\rangle = 0$ are shown by the dashed lines. In this case, the agreement with the data in the high-$p_T$ region becomes visibly poorer.
Thus, we can confirm the known
difficulties\cite{HQSSPuzzle-1, HQSSPuzzle-2} in the description of the first LHCb data on $\eta_c$ production within the NLO NRQCD.
However, it is
worth noting a substantial discrepancy between the cross sections reported by the two LHCb analyses at 13 TeV. These data were collected during different data-taking periods but with similar selection criteria. The only difference is in rapidity regions, namely, $2 < y < 4.5$ and $2 < y < 4$, which should very weakly affect the differential cross sections.
Anyway, the problem of accurate description of $\eta_c$ production at the LHC still awaits a solution.

\begin{figure}
\begin{center}
{\includegraphics[width=.49\textwidth]{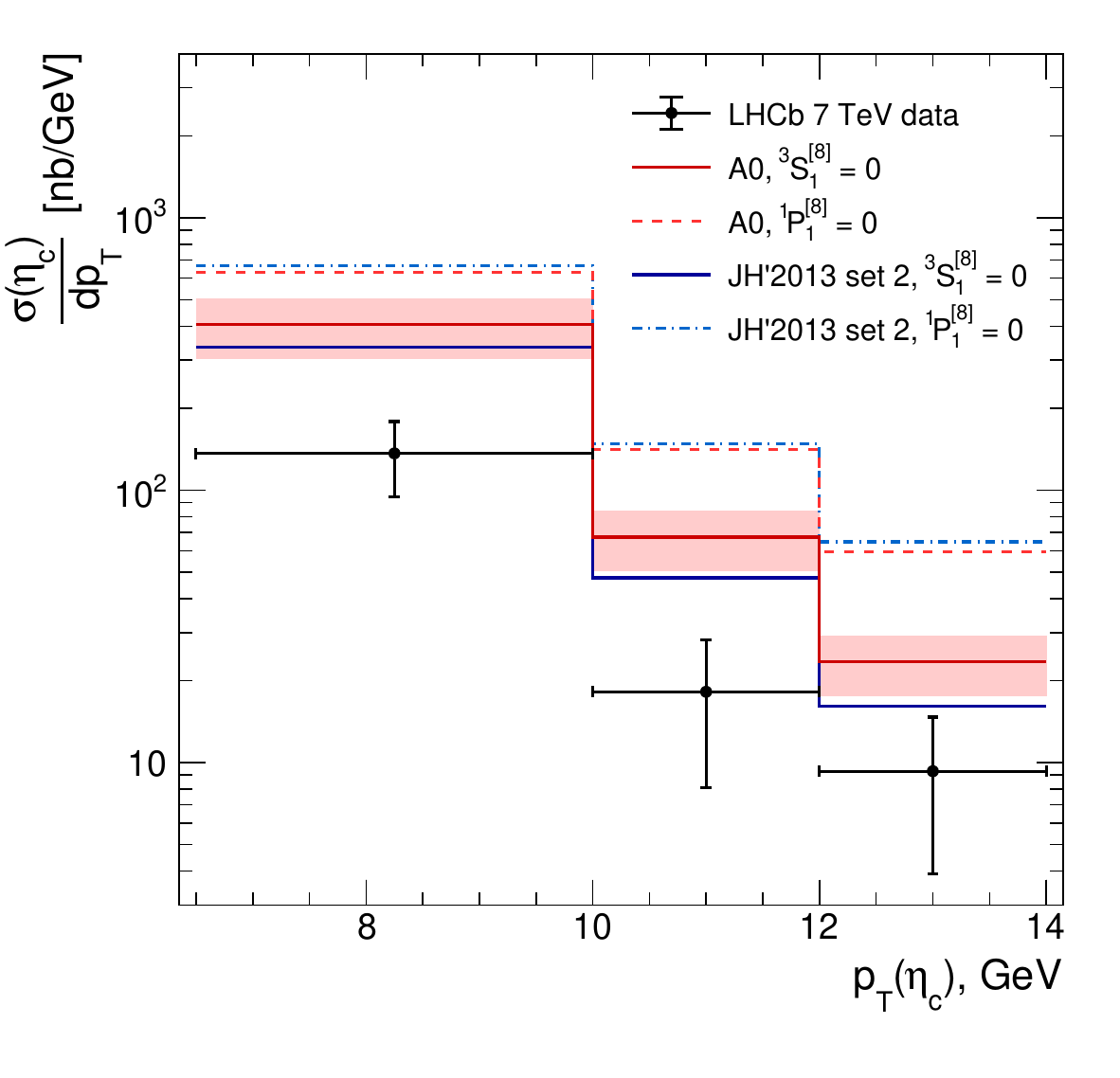}}\hfill
{\includegraphics[width=.49\textwidth]{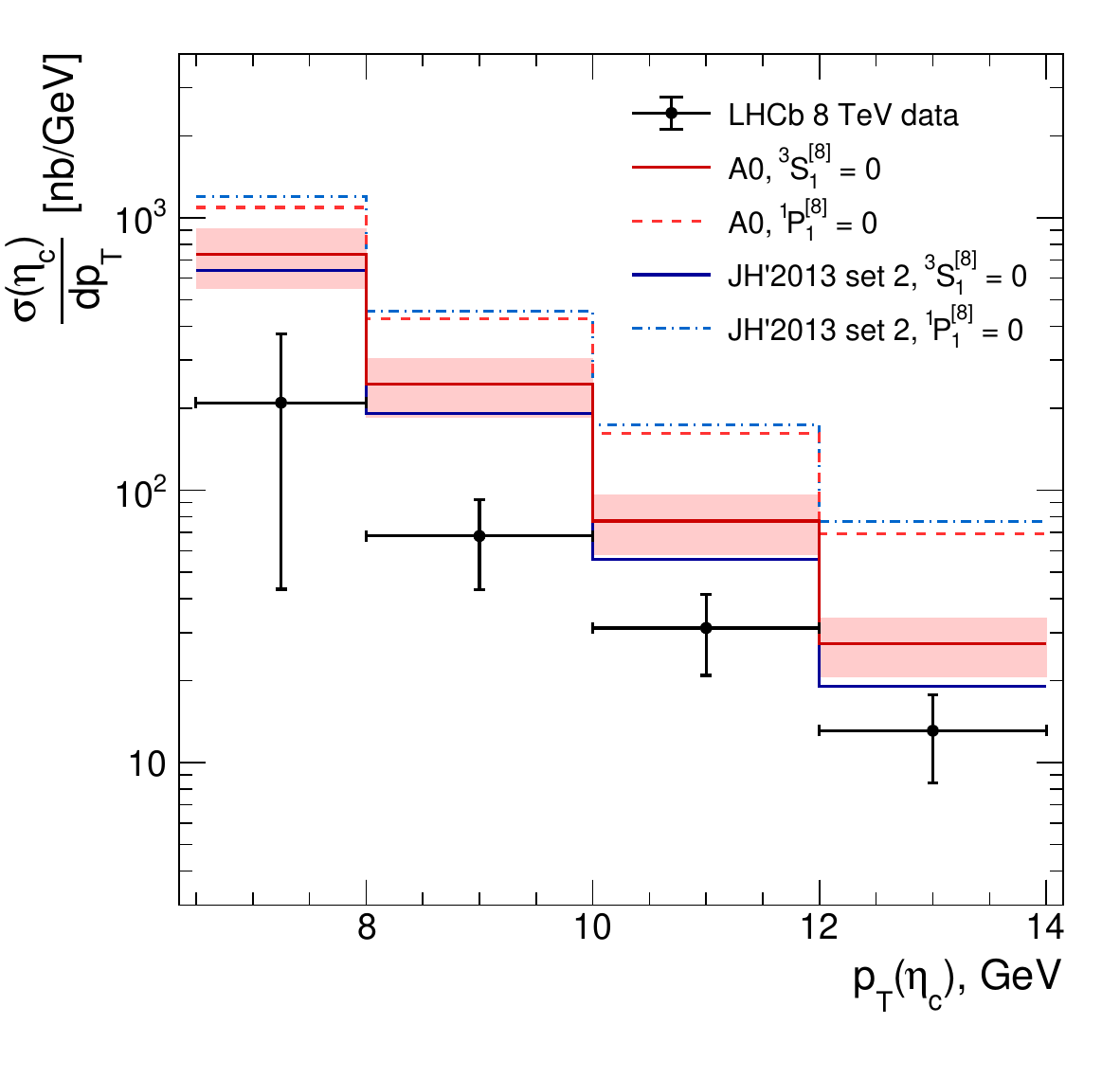}}\hfill
{\includegraphics[width=.49\textwidth]{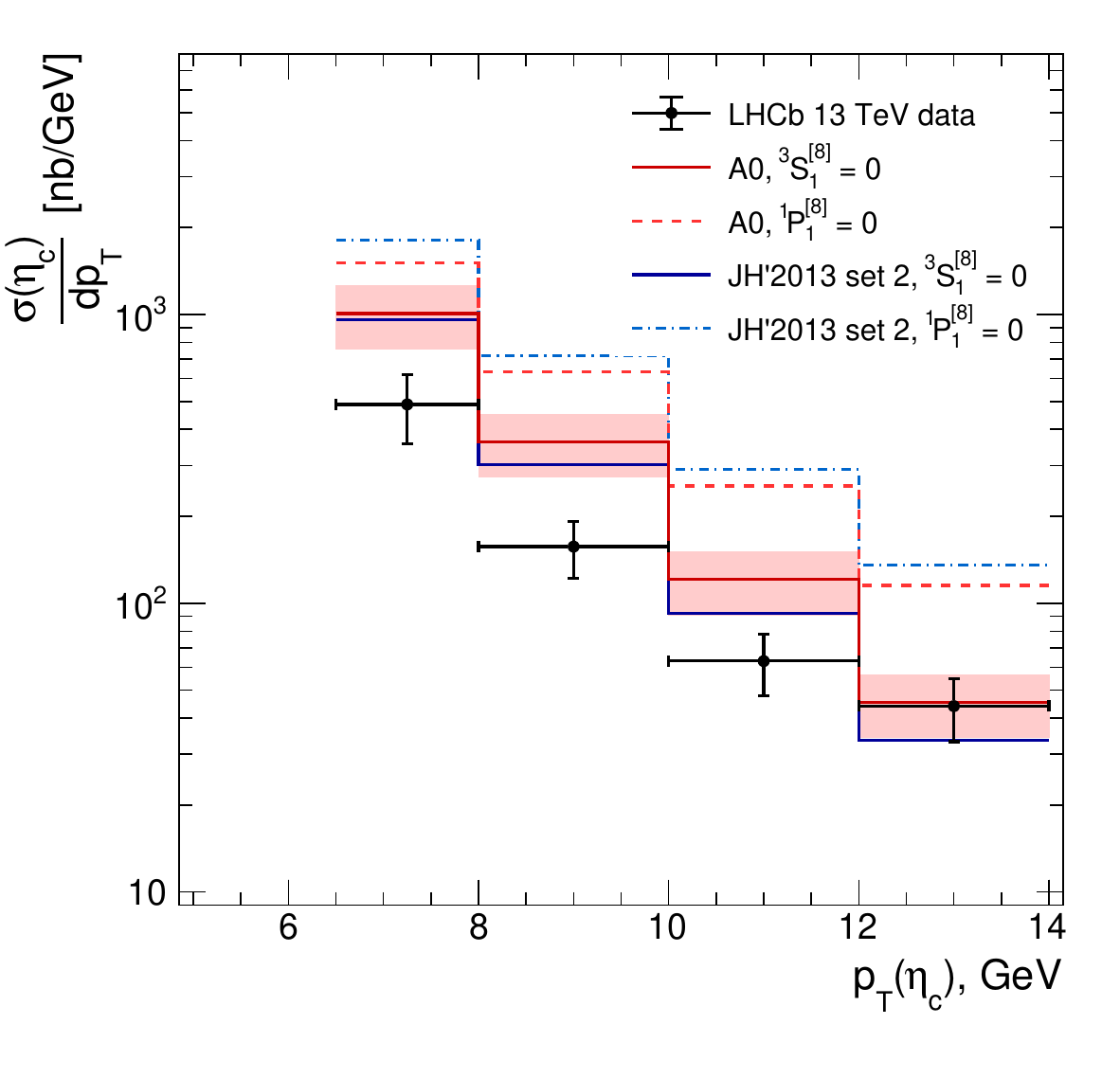}}\hfill
{\includegraphics[width=.49\textwidth]{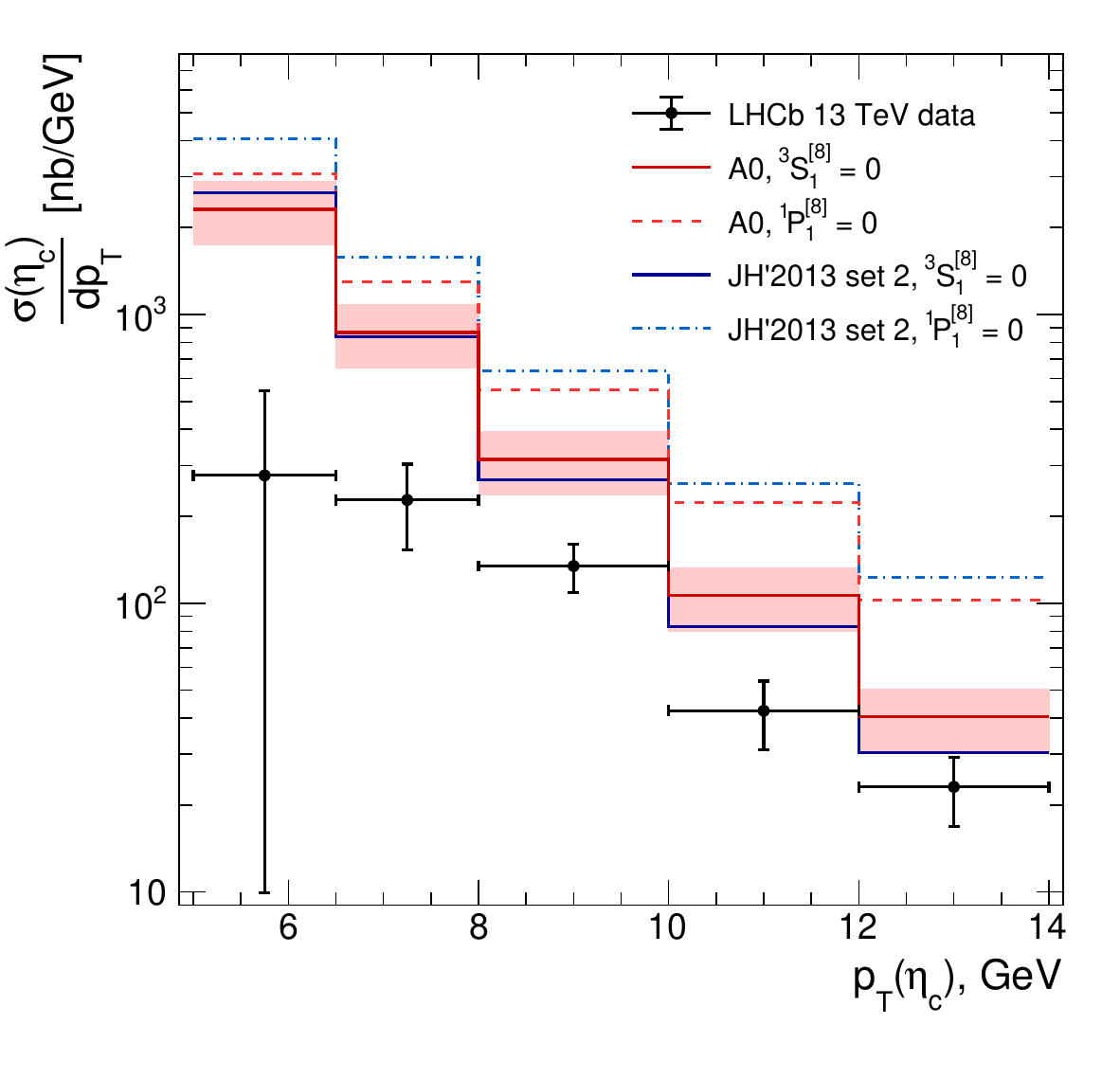}}\hfill
\caption{The differential cross sections for inclusive $\eta_{c}$ production in $pp$ collisions at $\sqrt{s} = 7, 8$ and $13$ TeV  as functions of the $\eta_{c}$ transverse momentum at the forward rapidity regions. The LHCb data are taken from \cite{dataset_etac_LHCb7_8, dataset_etac1_LHCb13, dataset_etac2_LHCb13}.}
\label{fig:LHCb_etac}
\end{center}
\end{figure}

\vskip 2mm

\section{Conclusions} \indent

In the present paper we have considered prompt $J/\psi$, $\chi_{cJ}$ and $\eta_{c}$
hadroproduction in the high energy factorization
beyond the usual leading order approximation.
For the first time we took into account tree-level next-to-leading
contributions using a dedicated consistent scheme to merge the LO and NLO
off-shell production amplitudes (NLO$^{\dag}$ scheme). This approach implies special kinematic
restrictions to avoid possible double counting.
Using two different CCFM-evolved gluon densities in the proton, we
have extracted long distance matrix elements for $J/\psi$ and $\chi_{cJ}$ mesons
from a combined fit to LHC data, which include the differential and
double-differential distributions in $p_T$ and $y$ and ratios of
different production cross sections.
A reasonable simultaneous description of all available unpolarized data
for $J/\psi$, $\psi^\prime$ and $\chi_{cJ}$ mesons,
including their different relative production rates,
is achieved in the whole kinematic region.
We demonstrated that taking into account tree-level NLO corrections
leads to an improved description of the data.

Our fit for $\chi_{cJ}$ mesons still points to some violation of HQSS relations, which could be
above the expected in NRQCD relativistic effects.
However, the NLO$^{\dag}$ corrections significantly reduce
such violations in comparison with the LO calculations. 

The predicted $\eta_c$ cross sections are above the available LHCb data.
However, we point out a substantial discrepancy between the corresponding transverse momentum
spectra reported in the two recent LHCb analyses.
Thus, the problem of accurate description of $\eta_c$ production at the LHC still awaits a solution.

\section*{Acknowledgements} \indent

We thank H.~Jung for his interest, very important comments and remarks.
The study has been supported by the Russian Science Foundation grant No.~25-22-00066, https://rscf.ru/en/project/25-22-00066/.

\bibliography{psi}

\end{document}